\documentclass[smallextended]{svjour3}       

\usepackage{graphicx}
\usepackage{mathptmx}      
\usepackage{aps-bibstyle}  
\journalname{Exp Astron}
\begin{document}

\title{A fast 2D image reconstruction algorithm from 1D data for the {\it Gaia} mission
\thanks{This work was supported by the STFC grant number RG59671.}
}

\author{D.L. Harrison  
}


\institute{D.L.Harrison \at
Institute of Astronomy \& Kavli Institute for Cosmology, University of Cambridge, Madingley Road, Cambridge, CB3 0HA, United Kingdom \\
              \email{dlh@ast.cam.ac.uk}           
}

\date{Received: date / Accepted: date}

\maketitle

\begin{abstract}
 A fast 2-dimensional image reconstruction method is presented, which takes as input 1-dimensional data acquired from scans across a central source in different orientations.  The resultant reconstructed images do not show artefacts due to non-uniform coverage in the orientations of the scans across the central source, and are successful in avoiding a high background due to contamination of the flux from the central source across the reconstructed image. Due to the weighting scheme employed this method is also naturally robust to hot pixels.  This method was developed specifically with {\it Gaia} data in mind, but should be useful in combining data with mismatched resolutions in different directions.

 \keywords{methods: data analysis}

\PACS{95.75.Mn \and 95.80.+p }


\end{abstract}

\section{Introduction}
\label{intro}

{\it Gaia} is an European Space Agency,  ESA, satellite mission, due for launch in 2013, which will be inserted into a Lissajous orbit around the L2 Lagrange point of the Sun-Earth system. It is an astrometric mission which will improve upon the previous ESA  astrometry mission, {\it Hipparcos},  with 10,000 times the number of stars observed and increasing the parallax and proper-motion accuracy attained by an order of magnitude, \cite{vanLeeuwen07}. The {\it Gaia} catalogue will amount to about one billion stars, or 1 per cent of the Galactic stellar population, complete to 20th magnitude. This catalogue will consist of positions, proper motions, parallaxes, radial velocities, as well as astrophysical information derived from the on-board multi-colour photometry.  This will allow the first 3-dimensional map of our galaxy, and enable studies of its composition, formation and evolution. {\it Gaia} will also detect tens of thousands of extra-solar planetary systems, survey huge numbers of solar system objects and observe extragalactic objects, such as quasars and supernovae.  {\it Gaia} will also provide a number of stringent new tests of general relativity and cosmology.  These and other science goals are described in \cite{mignard05}.

Whereas {\it Hipparcos} relied on photomultiplier tubes, a modulating grid, and a fixed pre-launch catalogue of objects to observe, {\it Gaia} will use charge-coupled devices, CCDs,  and will perform a full survey. In order reduce the data rate to a  manageable level only regions around detected sources will be  read out from the CCD and transmitted back to Earth. The area around the sources and the level of binning which will be performed  will depend on the magnitude of the detected source, with the data from fainter sources being reduced to 1-dimension. Even after this expedient  {\it Gaia} will still be producing on average 50 Gb of data every day and by the end of its 5 year mission will have amassed 100 Tb of data; further details of the {\it Gaia} mission may be found in \cite{lindegren08}. The processing of the {\it Gaia} data will hence provide many logistical and technical challenges, due to its volume and the complexity of the analysis required. An overview of the {\it Gaia} data processing is presented in \cite{mignard08}.

The spin rate of {\it Gaia} is 60 ${\rm arcsec.s}^{-1}$ and the axis maintains an angle of  $45^{\circ}$ to the Sun, while slowly precessing around the solar direction, completing a full revolution every 63 days.  After its 5-year mission each object will have been observed between 50 to 250 times depending on the position of the source, with the ecliptic latitude of the source  being the most important factor in determining the coverage.  The orientation of the focal plane as it transits the source will naturally vary over time, hence these multiple observations of the source may be used to produce a 2-dimensional image of the region surrounding the source from the 1-dimensional data of each transit. Indeed in order to achieve the full potential of the catalogue, this will be necessary both to increase the total number of sources, by detecting nearby objects in these images and hence allowing the necessary corrections to be made to the astrometric and photometric parameters of the primary source.

Any image reconstruction method which could be used systematically on  {\it Gaia} data will need to be fast, given the number of sources to which it will need to be applied, and robust to variations in the coverage, the point spread functions, cosmic-ray and solar proton hits. The feasibility and potential performance of two different  image reconstruction methods, in the specific case of {\it Gaia} data, have been previously investigated. The drizzle method, as described by \cite{fruchter02} was investigated by \cite{nurmi05},  and image reconstruction via an iterative least squares procedure regularised with a smoothing term was investigated in \cite{lindstroem06} and \cite{mary06}.  It is doubtful, however that either method  as described would be satisfactory in the case of the actual {\it Gaia} data set due to both the increase in the complexity of the actual data and the volume of the data processing required;  \cite{lindstroem06} and \cite{mary06}  acknowledged themselves that their method as presented was too computationally expensive to be feasible. This paper presents a method which may be up to the challenge of the actual {\it Gaia} data set.

Section~\ref{gaia_data}  gives a brief description of how {\it Gaia} will acquire data, and how the data is selected and binned prior to transmission back to Earth.  While this has been presented elsewhere, it is important to describe it here as it is necessary to understand the difficulties imposed on the image reconstruction by the data compression and windowing strategies.  Section~\ref{simu_data}  describes the simulations used in this paper.  The new method is described in Section~\ref{faststack}, after which a comparison with the drizzle method is presented in Section~\ref{drizzle}.

\section{Description of the {\it Gaia} data}
\label{gaia_data}

As described above not all the data is transmitted back to Earth, and for the fainter objects a significant level of binning, for the purposes of data compression, has been applied prior to transmission. In this section, the selection and transmission of the data in regions around detected objects is described in more detail.  It  will be necessary to define some terminology which will be used throughout this paper. A window refers to the region, around the detected object, on the CCD from which the data is acquired. A window is composed of a set of samples,  these samples are formed using the data in the CCD pixels, this may be directly with a one-to-one relation between samples and pixels, or pixels may be binned together to form the sample. The size of the window, and the number of pixels contributing to each sample depends on both the CCD as well as the magnitude of the object. 
\begin{figure}
\includegraphics[width=0.95\textwidth]{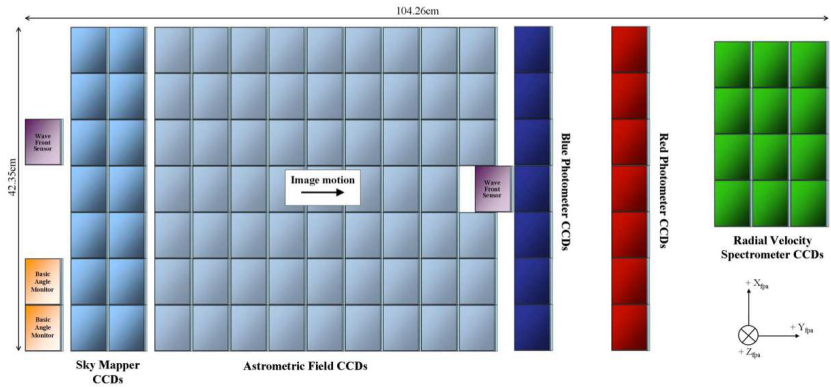}
\caption{The layout of the focal-plane of {\it Gaia}, showing the Sky Mapper, SM, and the Astrometric Field, AF. This image is courtesy of ESA and Alexander Short.}
\label{focalplane_fig}
\end{figure} The layout of the focal plane of {\it Gaia} is shown in Figure~\ref{focalplane_fig}. It consists of 106 large-format CCDs arranged over 7 rows and 17 strips. The largest section of CCDs forms the Astrometric Field, AF,  which is preceded by the Sky Mapper, SM,  which controls the object detection and selection of  window types in the subsequent CCDs transited by the object. The  Blue and Red Photometers  provide low resolution spectrophotometric measurements for each object and  the Radial-Velocity Spectrograph acquires  spectra of all objects brighter than about 17-th magnitude.

\begin{figure}
\includegraphics[width=0.95\textwidth]{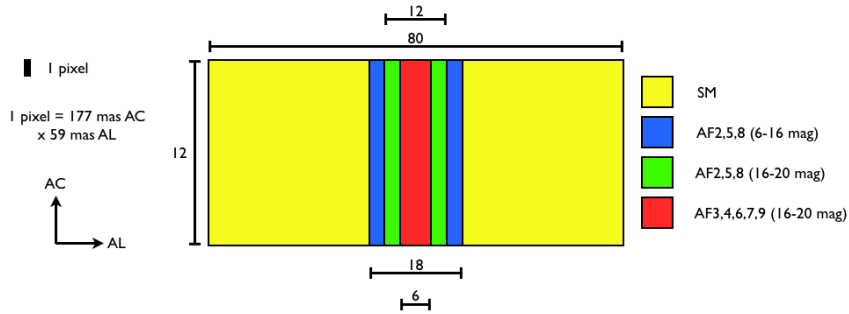}
\caption{This figure shows the relative sizes of the windows for objects of different magnitudes in different fields. It can be seen that the windows from the SM occupy the largest area, and this area remains the same for all magnitudes. The areas of the windows in the AF fields do, however, change with the magnitude of the object. For the brightest objects, magnitude 6-12, all the AF fields have the same area, shown in blue in the figure. For the fainter objects the size of the window also depends upon the strip in the AF, with strips 2,5 and 8 having larger windows. The AF2,5 and 8 windows are the same size for 12-16 magnitude objects as the 6-12 magnitude objects, only reducing in size for 16-20 magnitude objects. The relative size of a CCD pixel is also shown for comparison, it should be noted that the pixels are not square but elongated in the across scan, AC, direction with respect to the along scan, AL, direction.}
\label{window_footprint_fig}
\end{figure}

\begin{figure}
\includegraphics[width=0.95\textwidth]{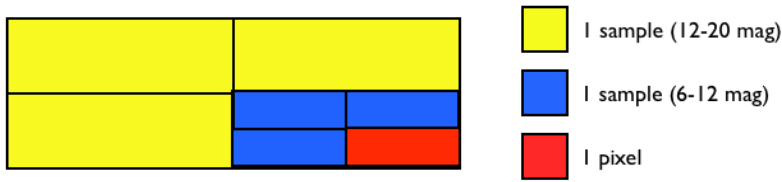}
\caption{While the overall size of the SM windows does not change with the magnitude of the observed object, the binning of the data does. This figure shows how the pixels are combined into the samples which are transmitted back to Earth. For windows centred on objects in the 6-12 magnitude range, a sample is formed from 4 pixels as shown by the blue sample; the windows are always read from the CCD in this format, but for fainter objects 4 blue samples are added together to create 1 yellow sample prior to transmission back to Earth.}
\label{sm_samples_fig}
\end{figure}

\begin{figure}
\includegraphics[width=0.75\textwidth]{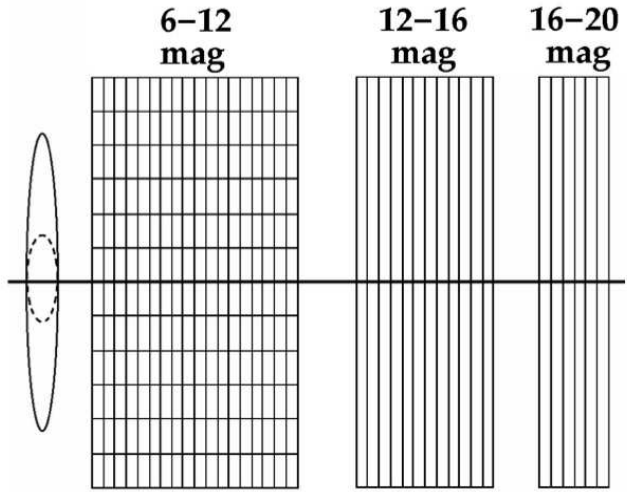}
\caption{The default astrometric field windows, used for AF strips 3,4,6,7 and 9,  are shown here. The dashed ellipse shows the ÔAiry diffraction discÕ of a single star; the solid ellipse shows the maximum smearing of the ÔAiry ellipseÕ in the across-scan direction, integrated over a single CCD transit, resulting from the precession of the spin axis.  The window on the left, for 6-12 magnitude objects, corresponds to the blue window area shown in Figure~\ref{window_footprint_fig},  the window in the centre corresponds to the green window area, and the window on the right the red area. The window for the brightest objects is not binned, so that  1 sample is equivalent to 1 pixel, this hold true for all AF strips. The window for the intermediate objects, 12-16 magnitude, is reduced in size in the along-scan direction for the strips 3.4,6,7 and 9, but remains the same size for strips 2,5 and 8.  All the pixels in the across-scan direction are binned together, so that the resultant window consists of samples only in the along-scan direction.  The size of the windows for the faintest objects, 16-20 magnitude, is reduced in the along-scan direction to 6 pixels for the  AF strips 3,4,6,7 and 9,  and 12 pixels for the  strips 2,5 and 8. This image is courtesy of ESA and Erik H\o g. }
\label{windowbinning_fig}
\end{figure} 

Figure~\ref{window_footprint_fig} illustrates the sizes of the windows for the SM and AF fields for the different magnitude ranges. All the windows are the same size in the across-scan, AC, direction, with the variation in the size coming from the along-scan, AL, direction. The SM windows are the largest and remain the same size for all magnitudes; however the number of pixels forming one sample changes for the fainter sources, as shown in Figure~\ref{sm_samples_fig}.  For the brightest sources the AF fields are all the same size, and each returned sample is equivalent to a CCD pixel, as shown in Figure~\ref{windowbinning_fig}. As the sources become fainter, however, the behaviour across the AF changes with some strips having long, referring to the AL direction, windows and others having short windows. The AF strips 2, 5 and 8 fall into the long-window category, the idea with the data produced by these strips is to capture information regarding the background; and it is envisaged that the implemented image reconstruction algorithm would limit the AF data used to these strips as the addition of the remaining strips would increase the computational overhead with little gain in terms of the reconstructed image. The AF windows for the intermediate, 12-16 magnitude, and faint sources, 16-20 magnitude, are essentially 1-dimensional as all the pixels in the AC direction are binned together to form a sample of size 12 AC pixels by 1 AL pixel, as illustrated in Figure~\ref{windowbinning_fig}, it is the number of these samples kept in the AL direction which differs between the intermediate and faint source windows with 18 (12) AL samples in the intermediate-magnitude source windows for AF strips 2,5 and 8 (3,4,6,7,and 9) and 12 (6) AL samples for the faint source windows.

\section{Simulated data}
\label{simu_data}

\begin{figure}
\includegraphics[width=0.95\textwidth]{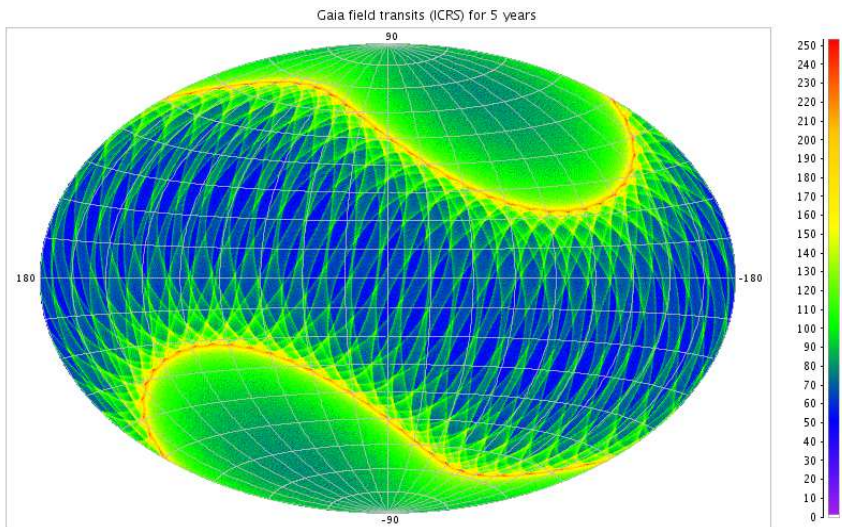}
\caption{The number of transits as a function of position on the sky at the end of 5 years, in equatorial coordinates.  This image is reproduced from the {\it Gaia} wiki ({\it http://www.rssd.esa.int/wikiSI/index.php?instance=Gaia} ) and is courtesy of Berry Holl.}
\label{transitsonsky_fig}
\end{figure}

\begin{figure}
\includegraphics[width=0.505\textwidth]{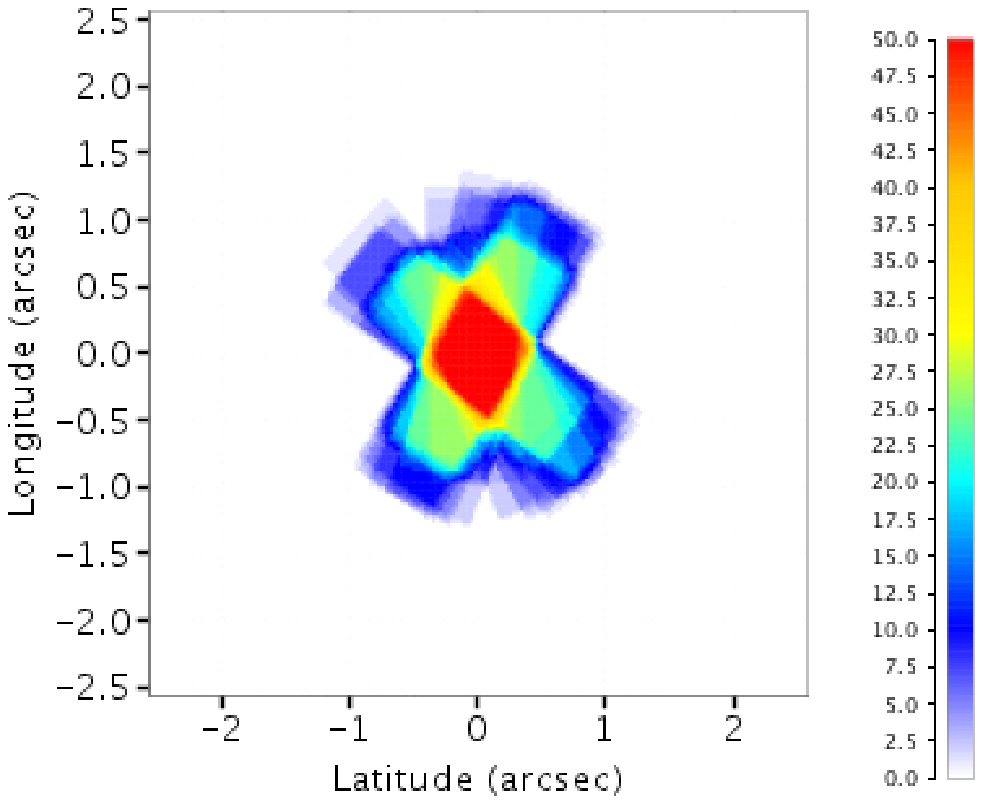}\includegraphics[width=0.505\textwidth]{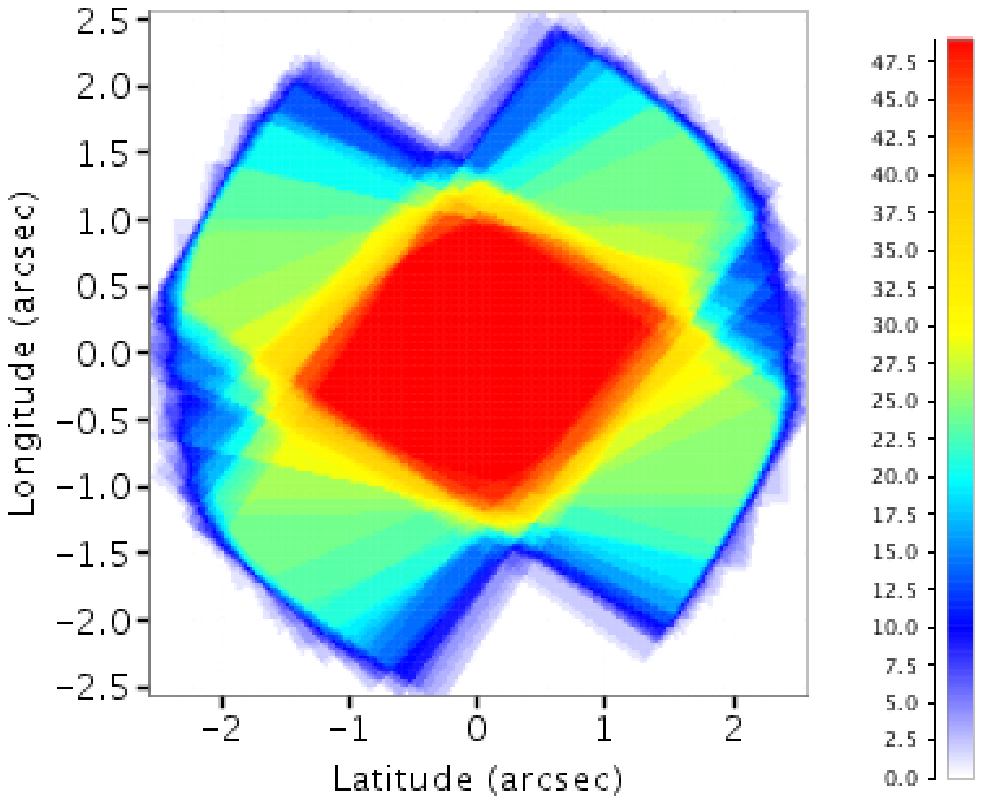}
\includegraphics[width=0.505\textwidth]{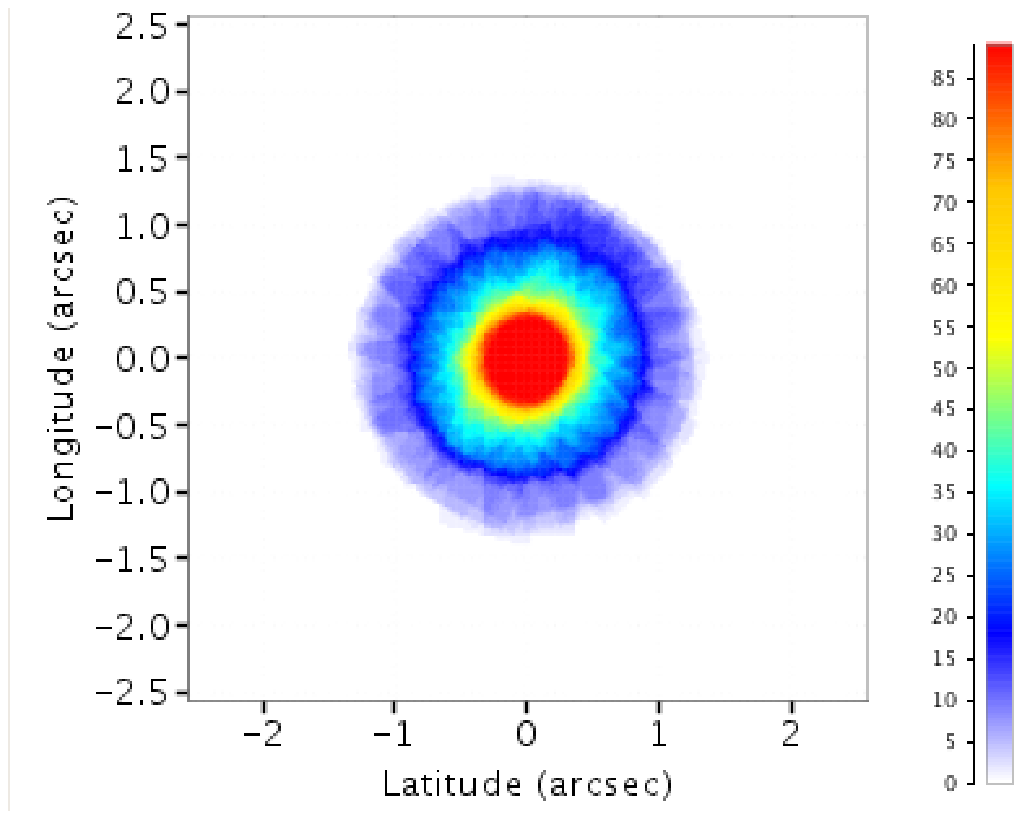}\includegraphics[width=0.505\textwidth]{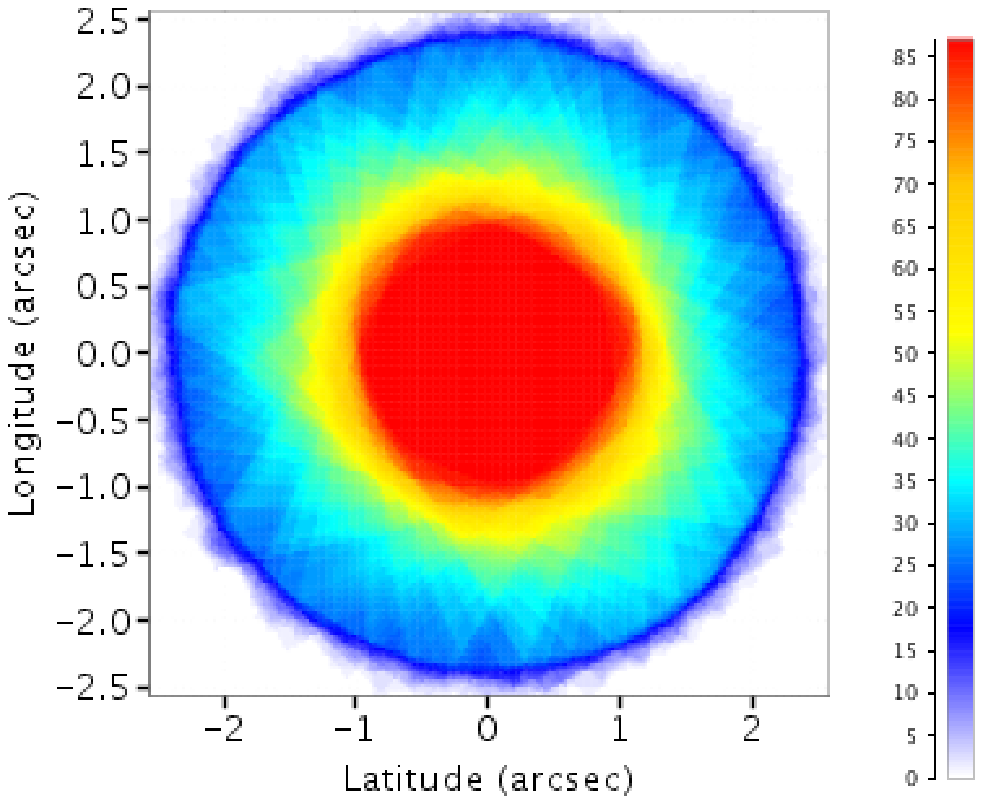}
\includegraphics[width=0.505\textwidth]{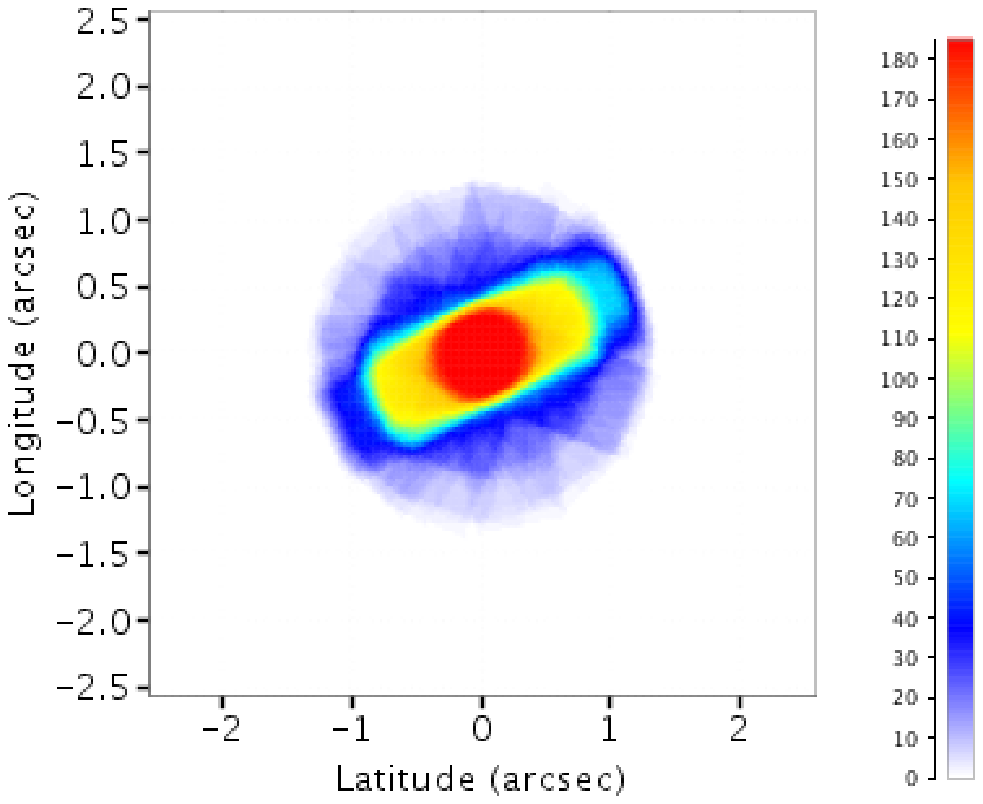}\includegraphics[width=0.505\textwidth]{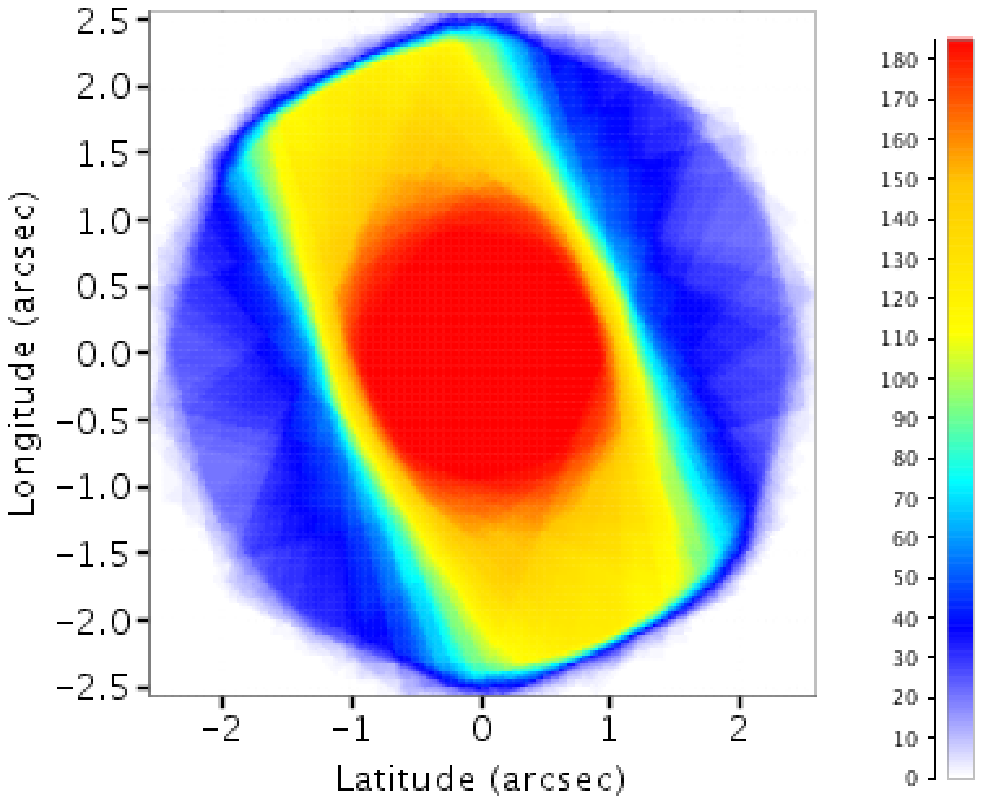}
\caption{This figure shows the coverage, in terms of contributing transits, in the vicinity of the primary due to the AF, for 12 AL pixel width windows, on the left and the SM on the right. The location of the primary source is different for each row of the figure, the coverage increases down the figure with the primary source being located in a low-coverage region in the top row, and average-coverage region in middle, and a high-coverage region at the bottom.}
\label{coverage_fig}
\end{figure}

The simulated data used for the image reconstructions in this paper made use of GIBIS,  Gaia Instrument and Basic Image Simulator,  as described by \cite{babusiaux05} and \cite{babusiaux10}. This allows the simulation of all the window data which would be observed by {\it Gaia} over the course of its 5-year mission for all sources located within a limited region of a given position on the sky. All locations on the sky are not equal in that the number, as shown in Figure~\ref{transitsonsky_fig}, and the distribution of the orientations of the transits varies over the sky.  As will be shown in section~\ref{drizzle}, it is important to test any candidate image reconstruction method at different locations as non-uniform coverage can produce artefacts in the reconstructed image. Figure~\ref{coverage_fig} shows the overlapping windows around a primary source located in regions with a low, average and high number of transits.  The distribution of the orientation of the transits in the region corresponding to the average number of transits is fairly uniform, whereas for the other two regions it is not. The AF and SM window coverages are shown separately in order to illustrate the full area around the source in which there is data, and the inner region in which there is higher resolution data, albeit in only one direction. 

\begin{table}
\caption{The locations of the secondary sources with respect to the primary, as shown in Figure~\ref{input_image_fig}. This is the same arrangement of secondaries that was used in \cite{lindstroem06} and \cite{mary06}. The sources B1, B2, and B3 lie within the area covered by the AF fields, the SM fields include all the sources up to and including B6, leaving B7 outside of the observed area. All the secondaries have the same magnitude in any one simulation, but as this magnitude varies over the different simulations used in this paper, no magnitudes are given in this table.}
\label{input_sources_table} 
\begin{tabular}{llll}
\hline\noalign{\smallskip}
relative & relative &  & distance \\
latitude & longitude & name & to primary  \\
(arcsec) & (arcsec) & & (arcsec) \\
\noalign{\smallskip}\hline\noalign{\smallskip}
-0.24 & 0.18 & B1 & 0.29 \\
0.59 & 0.00 & B2 & 0.59 \\
-0.59 & -0.59  & B3 & 0.83 \\
0.88 & 1.18 & B4 & 1.47 \\
-1.77 & 0.00 & B5 & 1.77 \\
2.36 & 0.00 & B6 & 2.36 \\
2.36 & -1.77 & B7 & 2.95 \\
\noalign{\smallskip}\hline
\end{tabular}
\end{table}

\begin{figure}
\includegraphics[width=0.505\textwidth]{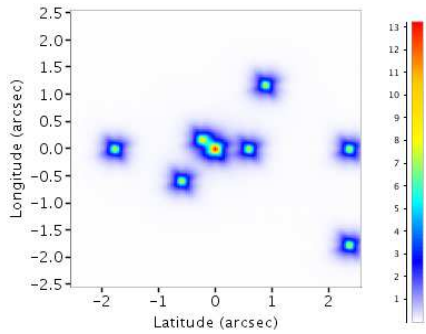}
\caption{This figure shows the sky in the vicinity of the primary located at (0,0) with the point sources convolved with the effective PSF, accounting for the coverage of the sky in different orientations. This is the same layout that was used in \cite{lindstroem06} and \cite{mary06}, in order to  aid the reader who wishes to compare the results with those obtained there.}
\label{input_image_fig}
\end{figure}

\begin{figure}
\includegraphics[width=0.72\textwidth]{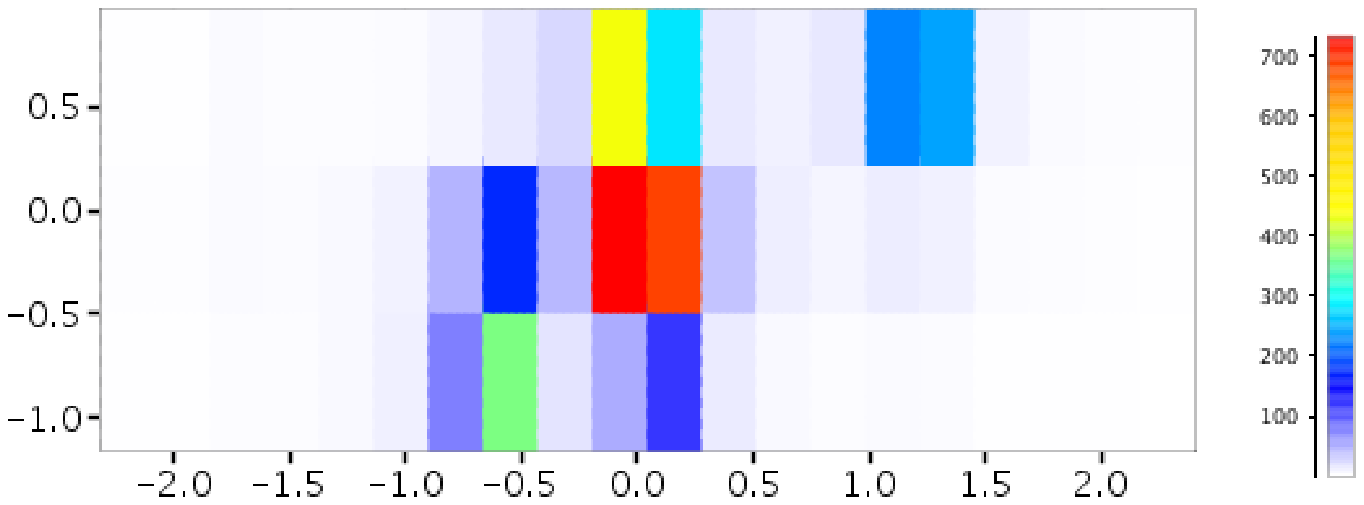}\includegraphics[width=0.22\textwidth]{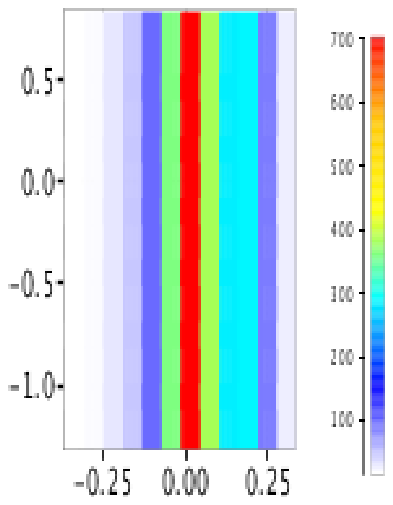}
\caption{This figure shows simulated SM and AF2 windows, which were produced by windows centred on the primary source, with the secondary sources laid out as shown in Figure~\ref{input_image_fig}.  All the secondary sources have the same magnitude, which is half a magnitude fainter than the primary magnitude which is 18. The orientation of the transit is such that  the along-scan direction is aligned left-to-right across the image of Figure~\ref{input_image_fig}.}
\label{simu_window_fig}
\end{figure}

The same relative layout of the secondary sources around the primary is used throughout this paper, with three different absolute locations used for the primary corresponding to the three different coverages shown in Figure~\ref{coverage_fig}.  This layout is illustrated in Figure~\ref{input_image_fig} in which the secondaries are all half a magnitude fainter than the primary. Figure~\ref{input_image_fig} is in essence the effective point spread function, PSF, convolved with the point sources, where the effective PSF is formed from the PSF as averaged through all the orientations in which the region is transited. In other words the image in Figure~\ref{input_image_fig} shows what could be achieved through stacking the individual transits, if the full CCD data were transmitted back to Earth. The relative locations of the secondaries with respected to the primary are also shown in Table~\ref{input_sources_table}; no magnitudes are given as these are allowed to vary, though the secondaries always all have the same magnitude as each other in any one simulation and the magnitude difference between the primary and the secondaries is always stated. To illustrate the appearance of  the actual data after the windowing and binning has occurred, an SM and an AF2 window are shown in Figure~\ref{simu_window_fig}. To produce Figure~\ref{simu_window_fig} a transit orientation must be assumed, the windows shown are those that would be formed if the along-scan direction was aligned left-to-right across Figure~\ref{input_image_fig}

All the simulated data used in this paper is generated for a primary source with a magnitude of 18; hence all  windows are of an appropriate size and binning for a source of this magnitude. All the transit data is of good signal-to-noise and the dominant problem in the image reconstruction is the localisation of the flux in the 1-dimensional samples. Poisson noise has been included in these simulations, but makes little difference to the reconstructed images. 

\section{Faststack}
\label{faststack}

It is apparent that any image reconstruction method which will cope well with the 1-dimensional window-data, will need to be able to reject the contribution of the long, narrow window-samples from regions of the image to which it does not belong. Consider for example the source, B3, the flux recorded at its location per observation will depend on the orientation of the transit, as if the transit is such that B3 is aligned with the primary source in the across-scan direction the window sample containing the flux from B3 will be completely dominated by the primary source. On a transit perpendicular to this one, however, there will be minimal contamination from the primary. If one were to collect all the contributions to a given pixel in the reconstructed image, then it stands to reason that the lower values are more likely to be a true representation of the flux at that location rather than the higher values, as these are likely to be due to window samples dominated by the primary source.  One could then consider a weighting scheme which penalised the higher values, for reconstructed image pixels with window samples which span a large range of values, as this behaviour is indicative of occasional contamination by the primary source.

The pixel in the $i^{th}$ row and the $j^{th}$ column of the reconstructed image, $r_{ij}$,  is then evaluated by:
\begin{equation}
r_{ij} = \frac{\sum^{N_w}_k \theta_k  w_k}{\sum^{N_w}_k w_k}
\label{recon_imagepix_eqn}
\end{equation} where $\theta$ is the array of window samples, $s_1, s_2, s_3 ... s_k ... s_{N_w}$ which contribute to this pixel.  A window sample contributes to a pixel in the reconstructed image, if the coordinates of the centre of the pixel lie within the area of the window sample. The window samples array,  $\theta$ is ordered in terms of increasing flux, such that $s_1 < s_2 < s_3 < s_k < s_{N_w}$ ,  where $N_w$ is the total number of window samples which contribute to the ${ij}^{th}$ pixel, and $w_k$ is the weighting applied to the $k^{th}$ window sample and is evaluated using equation~\ref{weight_eqn}. 

\begin{equation}
w_k = \left\{ \begin{array}{ll}
	\frac{1}{n_{s_k}}  & \mbox{$ k \leq \alpha N_w$} \\
	\frac{1}{n_{s_k}} \exp \left(1 - \left( \frac{\theta_k}{r_{ij_0}} \right)^2 \right)  & \mbox{$ k > \alpha N_w$}
       \end{array} \right.
\label{weight_eqn}
\end{equation} where $n_{s_k}$ is the number of pixels in the reconstructed image to which  the window sample, $s_k$, contributes, $\alpha$ is chosen such that $ 0 < \alpha < 1 $, and where  $r_{ij_0}$ is given by:
\begin{equation}
r_{ij_0} = \frac{\sum^{\alpha N_w}_k \theta_k  w_k}{\sum^{\alpha N_w}_k w_k}
\label{rzero_eqn}
\end{equation} 

The variance of the reconstructed image pixel, $r_{ij}$, may then be found using the standard equation for the sample variance of a weighted mean:
\begin{equation}
\sigma^2_{r_{ij}} = \frac{\sum^{N_w}_k  w_k \left(  \theta_k  - r_{ij}  \right)^2}{\sum^{N_w}_k  w_k}
\label{var_recon_imagepix_eqn}
\end{equation}

To see how this weighting scheme works, one may consider, $r_{ij_0}$, to be the first estimate for the $r_{ij}$ pixel value, which is formed from the window samples,  which have values which lie in the lowest part of the distribution, as determined by the $\alpha$ parameter. Returning to the example of a pixel located on the source, B3, these values would be ones in which the transit was aligned such that this pixel was not contaminated by flux from the primary.  The remaining window samples would enter into the assessment, with an additional factor in the evaluation of the weighting, designed to rapidly reduce the contribution of samples whose values are greatly in excess of the initial estimate.

\begin{figure}
\includegraphics[width=0.505\textwidth]{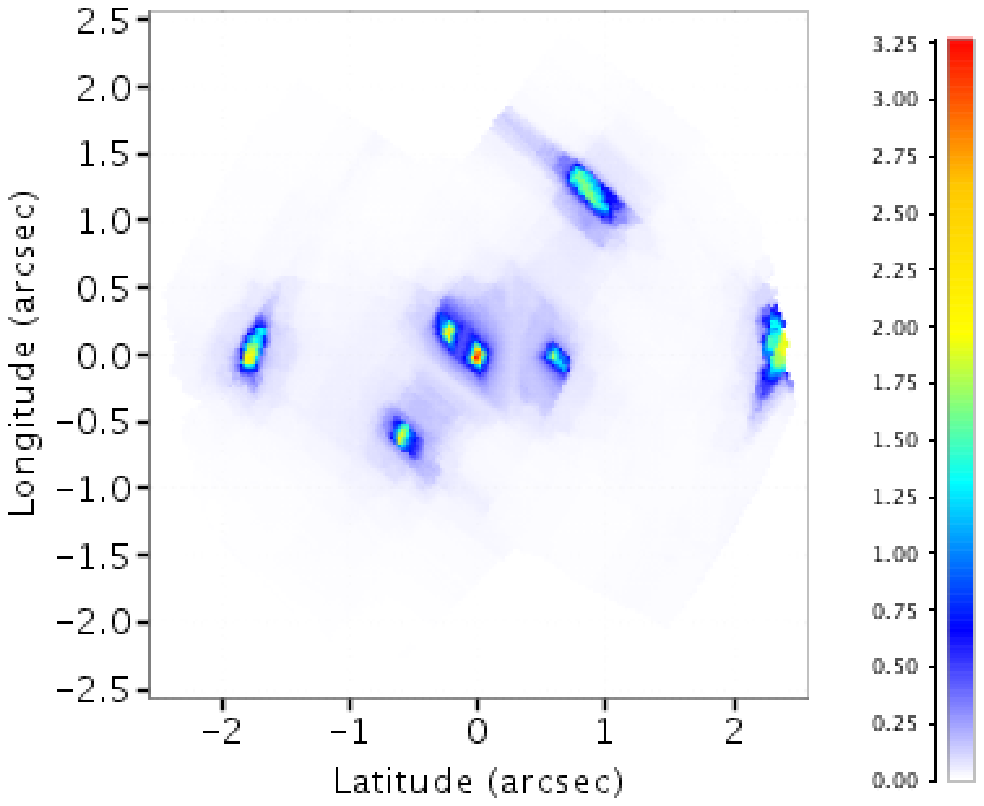}\includegraphics[width=0.505\textwidth]{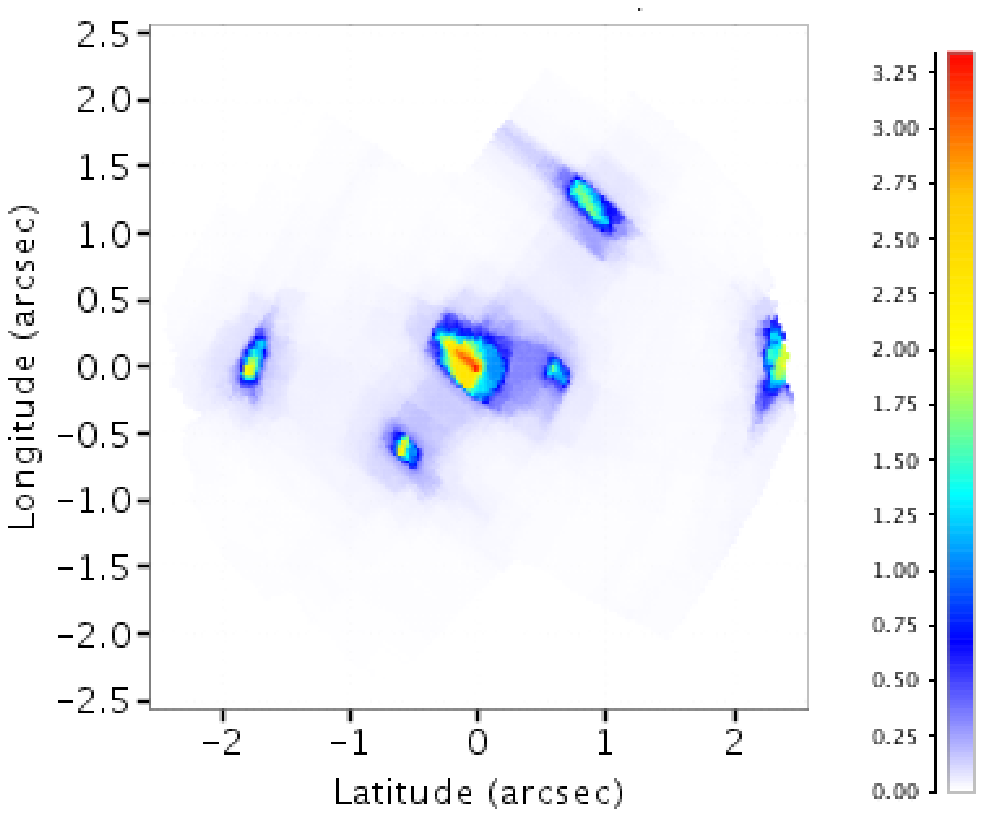}
\includegraphics[width=0.505\textwidth]{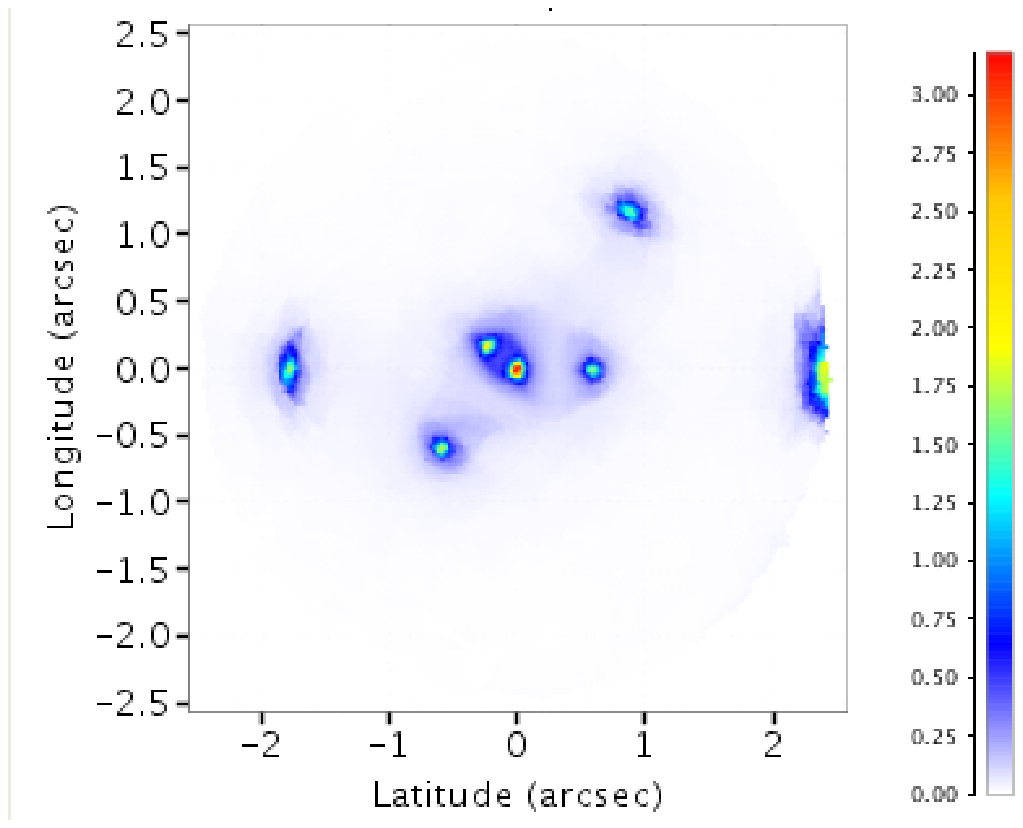}\includegraphics[width=0.505\textwidth]{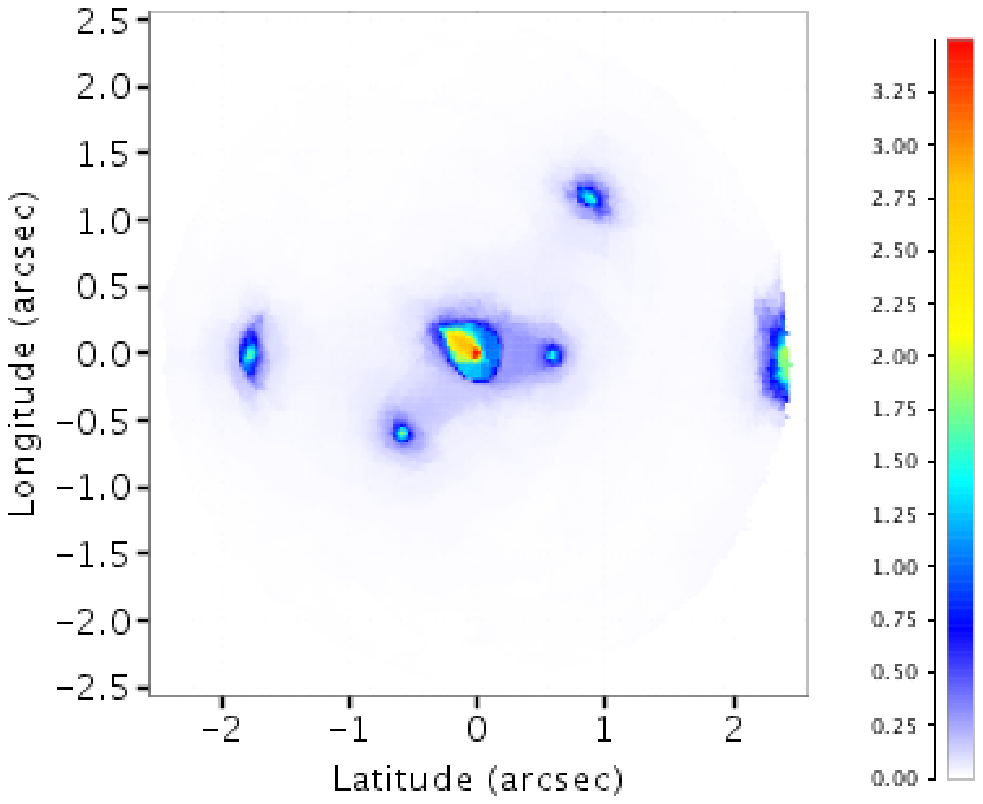}
\includegraphics[width=0.505\textwidth]{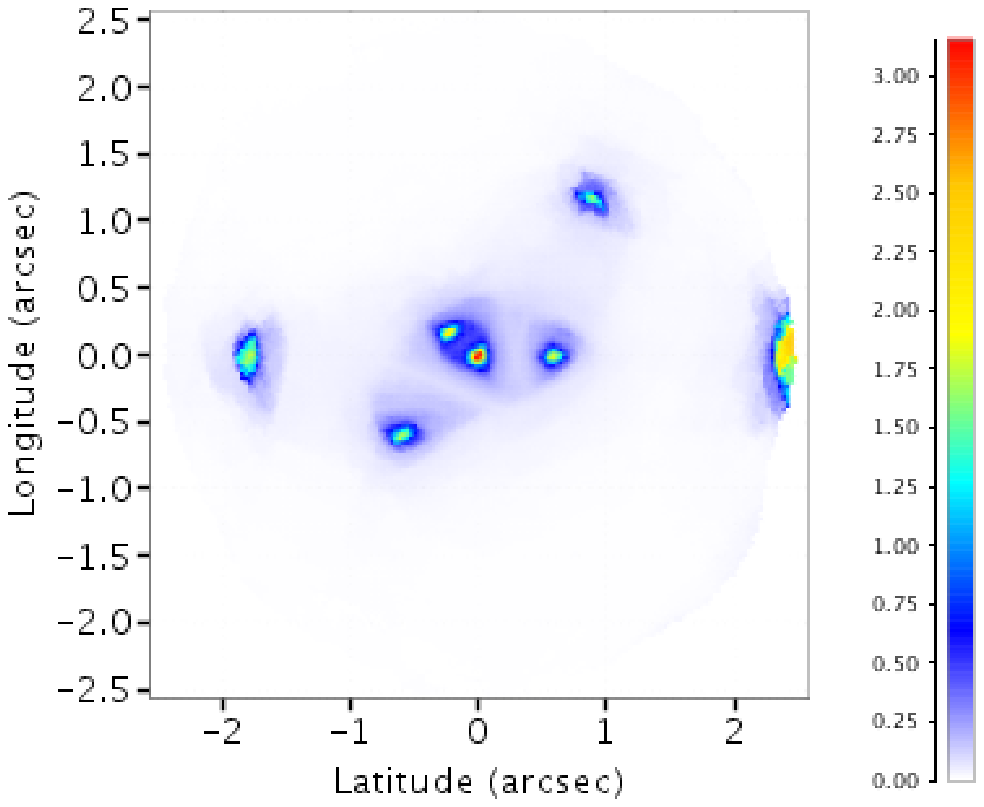}\includegraphics[width=0.505\textwidth]{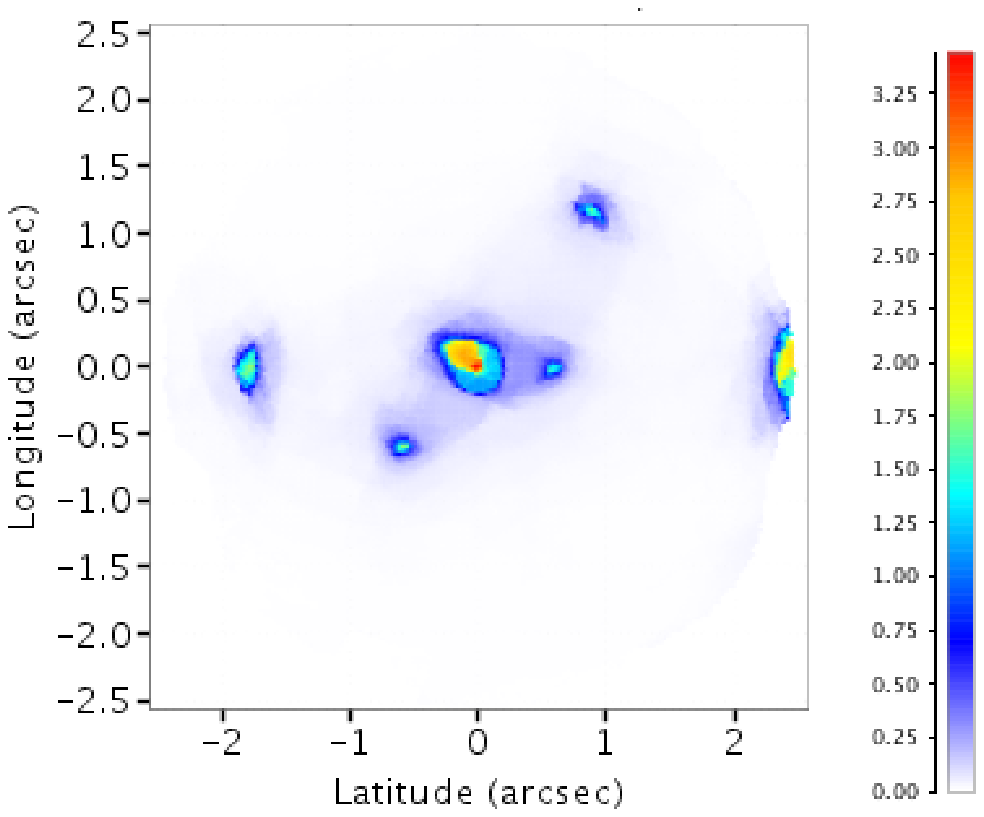}
\caption{This figure shows the images reconstructed by the faststack method in the regions corresponding to the coverages shown in Figure~\ref{coverage_fig}. The reconstructions of the left use the SM, AF2, 5 and 8 data, whereas the reconstructions on the right just use the SM data. This shows the increase in resolution possible when incorporating the AF data.}
\label{faststack_coverage_fig}
\end{figure}

\begin{figure}
\includegraphics[width=0.505\textwidth]{faststack_a3_lowalpha}\includegraphics[width=0.505\textwidth]{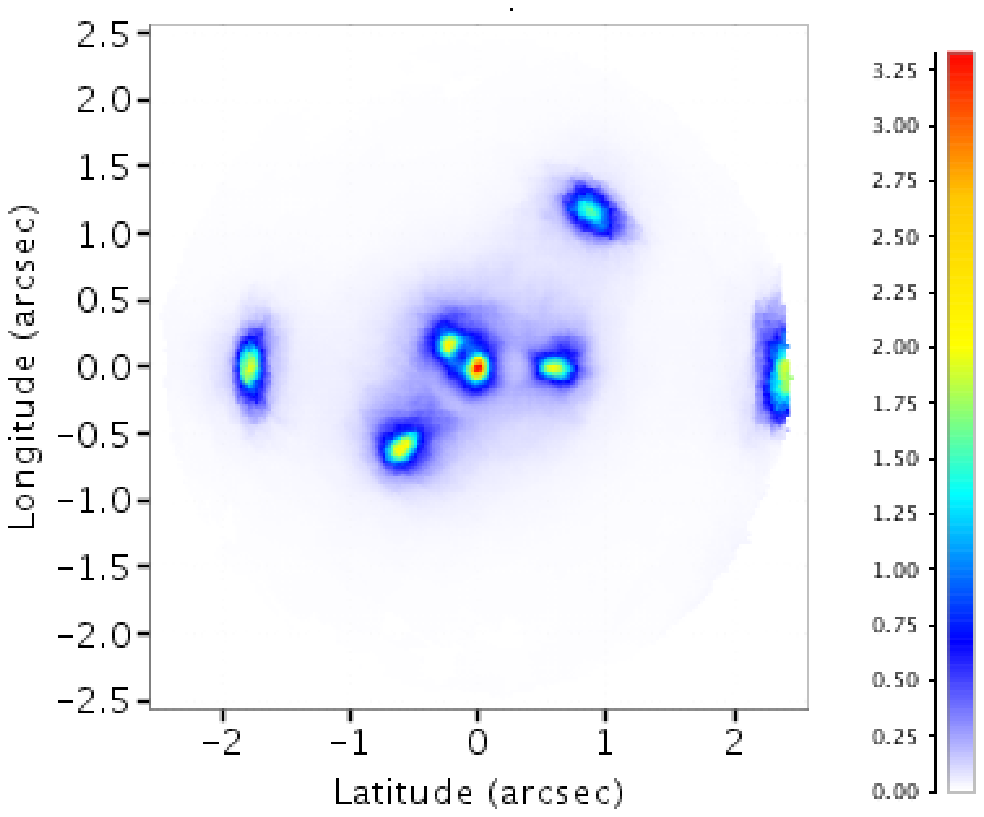}
\includegraphics[width=0.505\textwidth]{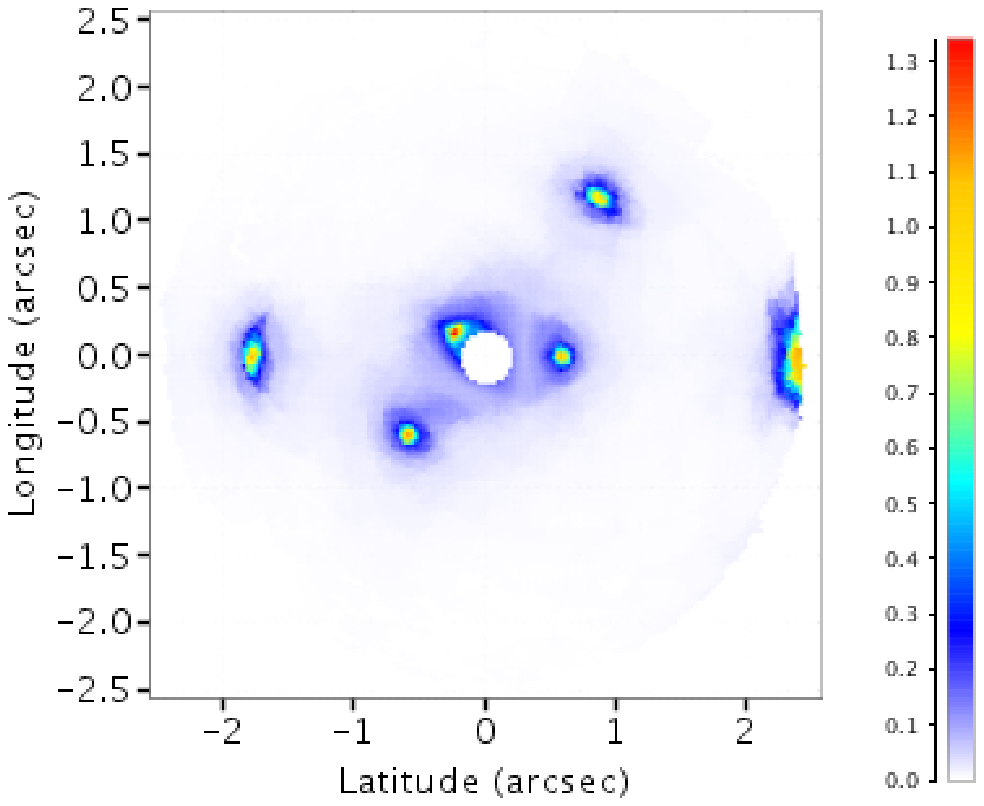}\includegraphics[width=0.505\textwidth]{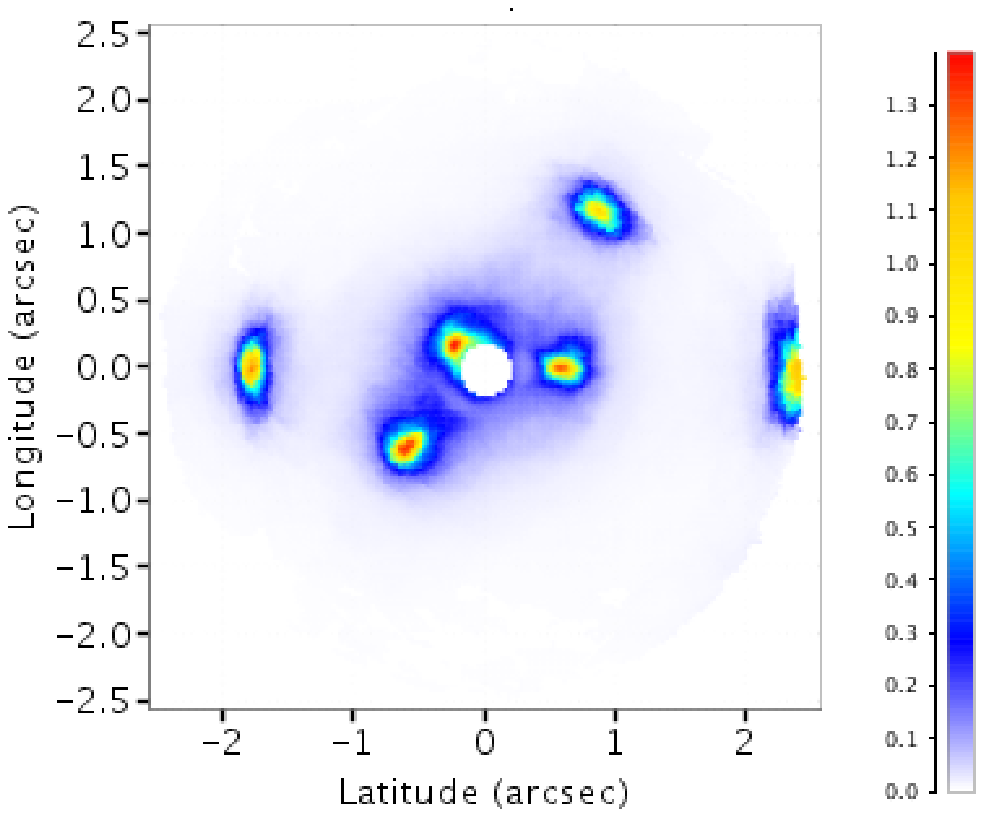}
\includegraphics[width=0.505\textwidth]{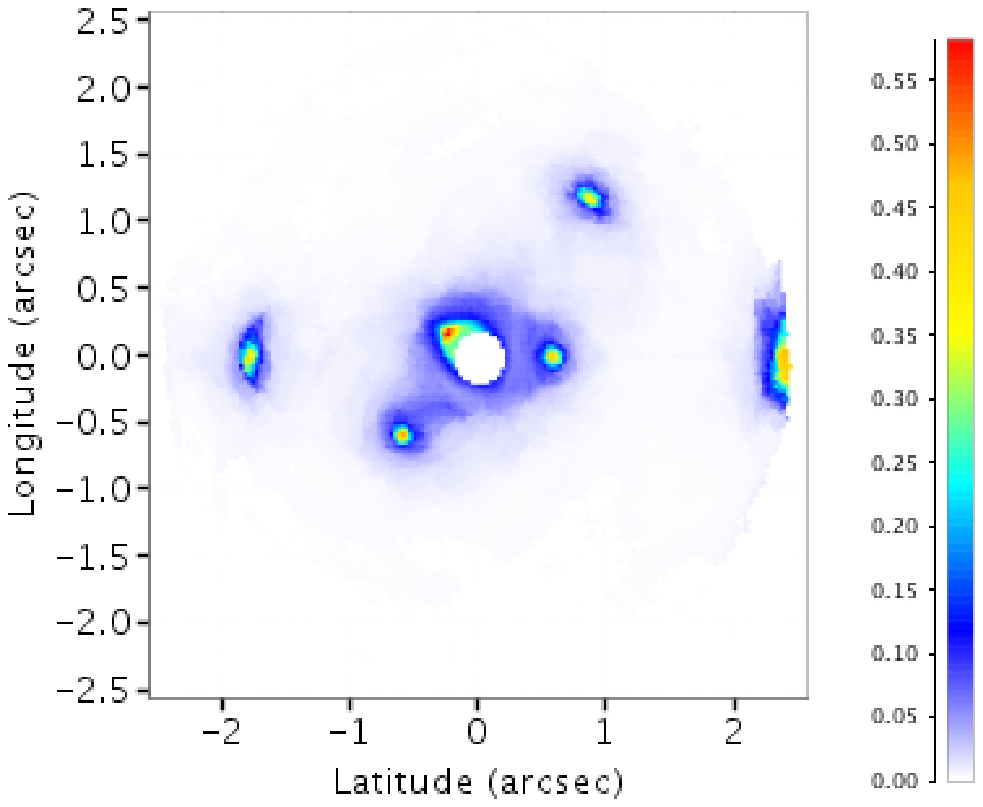}\includegraphics[width=0.505\textwidth]{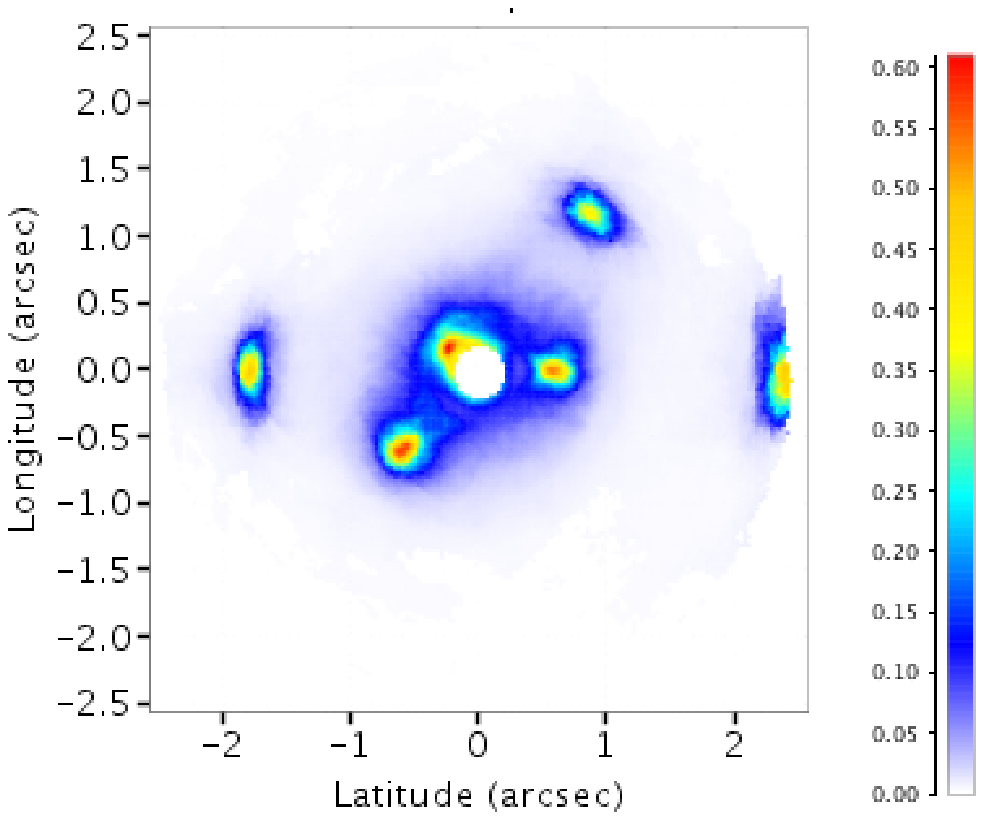}
\caption{This figure shows the faststack reconstructions, in the average coverage region, on the left for $\alpha=0.05$ and $\alpha=0.45$ on the right. The magnitude of the primary is the same throughout, but the magnitude of the secondary sources, and hence the magnitude difference from the primary, increases down the figure, and in its continuation on a subsequent page. The top row shows the reconstruction for the average coverage and a magnitude difference of a half. The second row shows the reconstruction after the magnitude difference has been increased to one; a $0.2^{\prime \prime}$ radius disc centred on the primary has been masked out, hence reducing the range of the colour scale in order to aid in the visualisation of the secondaries. The magnitude difference between the primary and the secondaries is increased by one for each subsequent row of the figure, reaching a magnitude difference of five by the last row of the continuation of this figure. }
\label{alpha_comparison_fig}
\end{figure}

\begin{figure}
\includegraphics[width=0.505\textwidth]{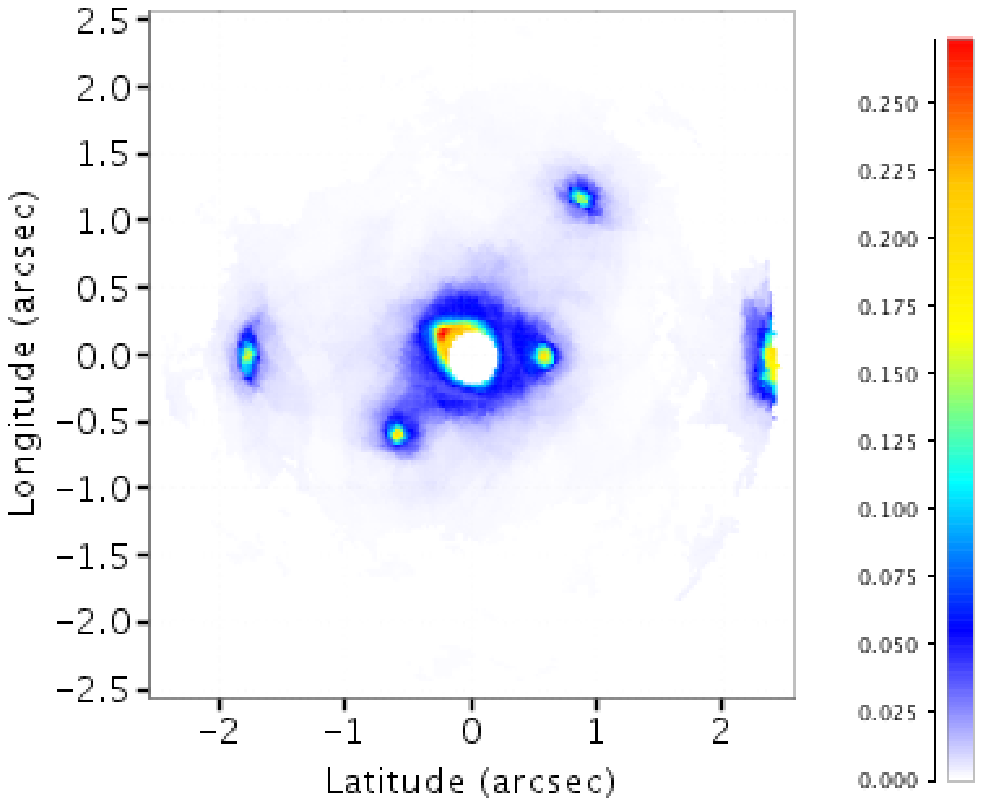}\includegraphics[width=0.505\textwidth]{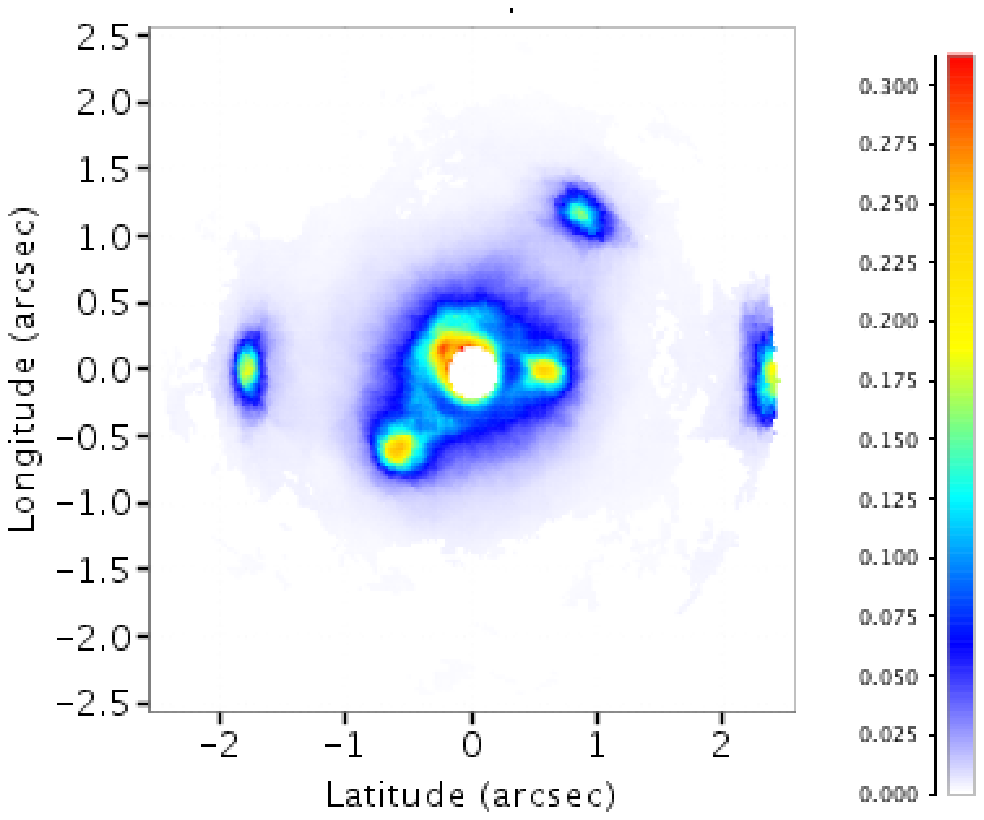}
\includegraphics[width=0.505\textwidth]{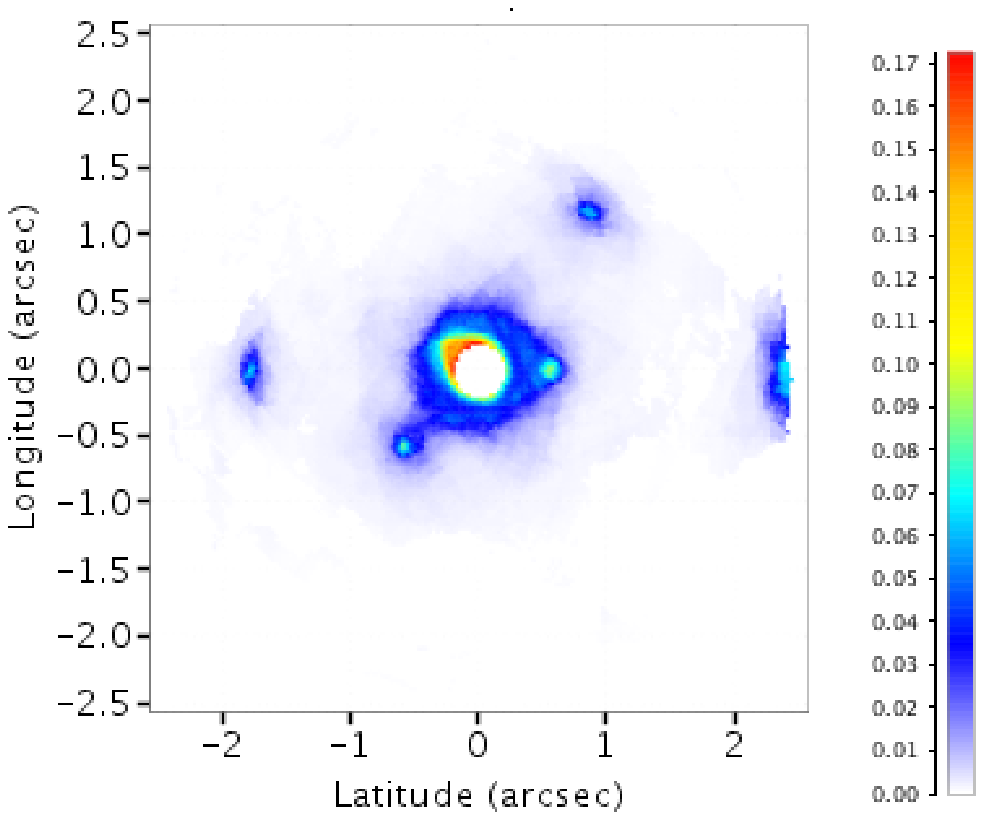}\includegraphics[width=0.505\textwidth]{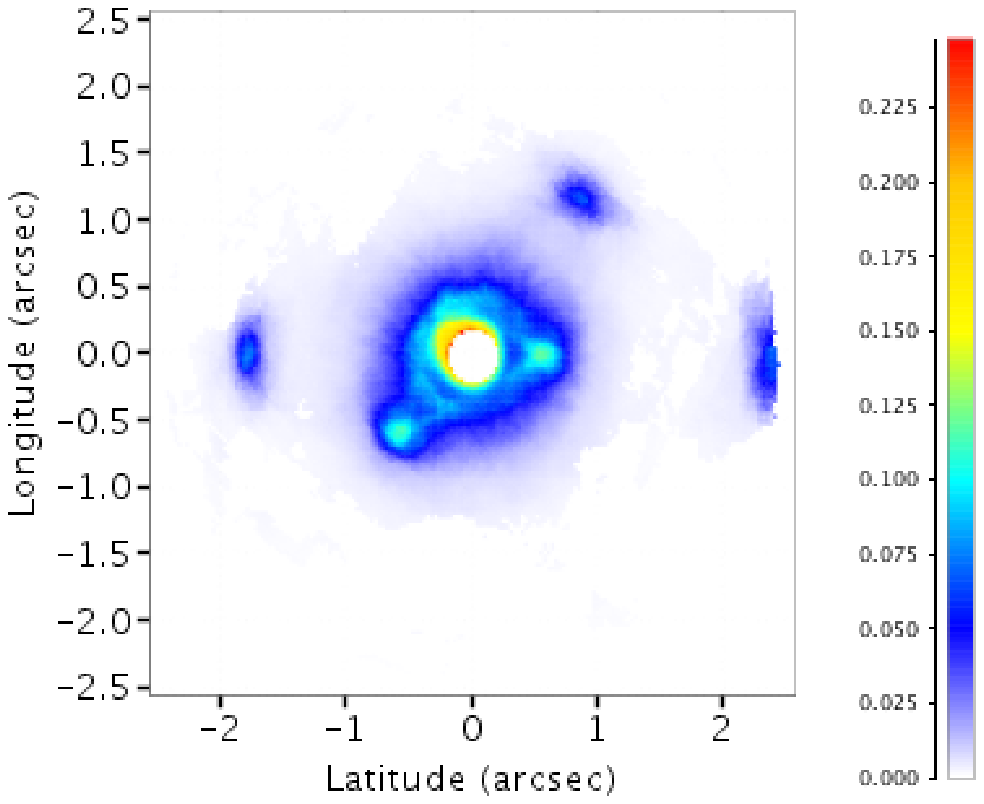}
\includegraphics[width=0.505\textwidth]{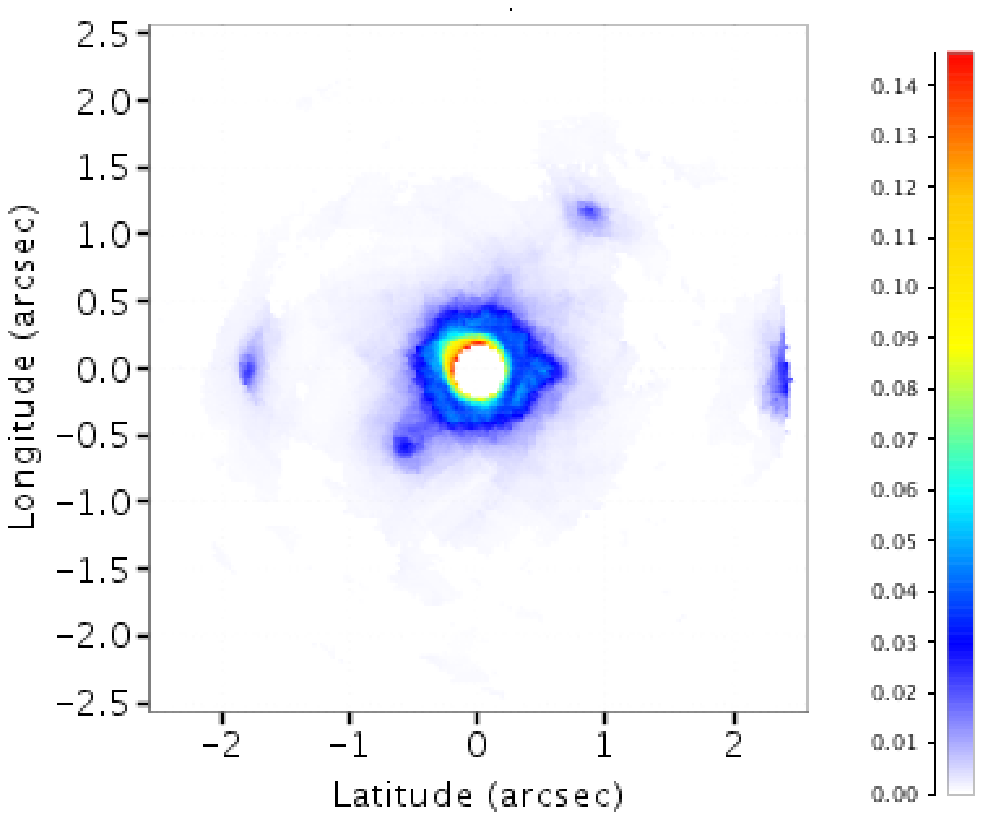}\includegraphics[width=0.505\textwidth]{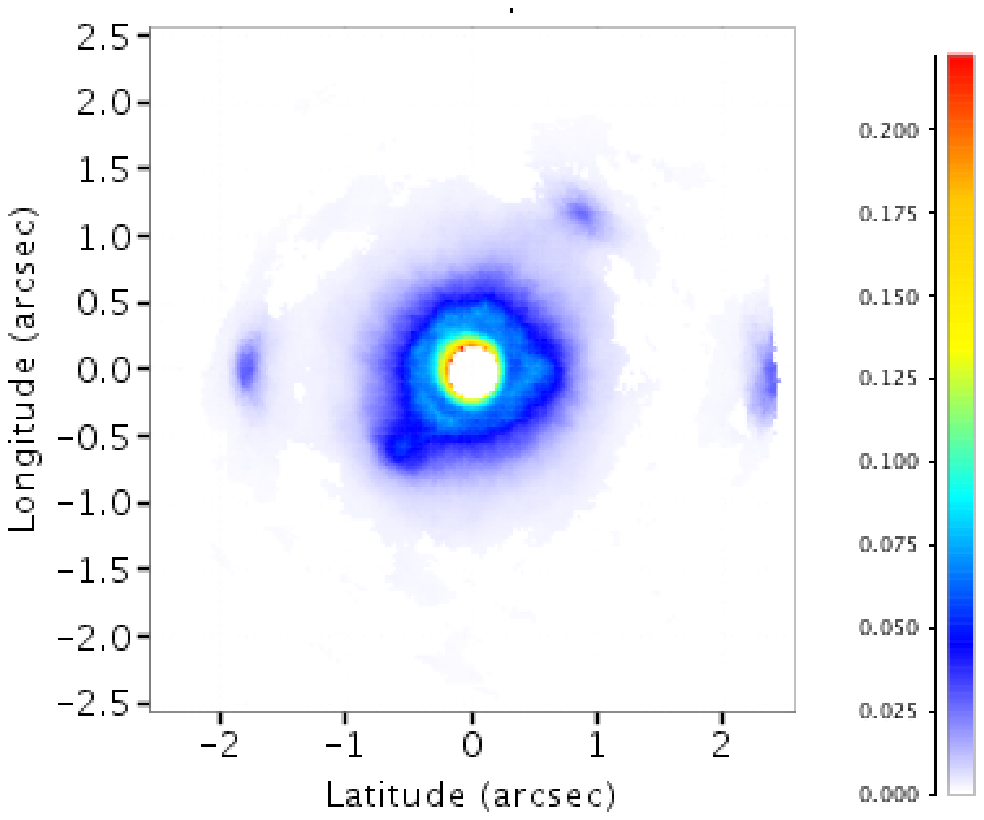}
Figure~\ref{alpha_comparison_fig} (continued)
\end{figure}

Figure~\ref{faststack_coverage_fig} shows the reconstructions made using the data simulated in each of the coverage regions shown in Figure~\ref{coverage_fig}, using the layout of sources as shown in Figure~\ref{input_image_fig} and Table~\ref{input_sources_table}. The data were produced for a primary magnitude of 18, meaning faint source windows,  and all the secondaries had a magnitude of 18.5. The reconstructions were made using the SM and AF2, 5 and 8 data, shown in the left column of Figure~\ref{faststack_coverage_fig}, and with just the SM data, shown in the right column.  Figure~\ref{faststack_coverage_fig}  shows the advantage of including the AF data in that the source B1 is detectable, whereas when just the SM data are used it is too close to the primary to be resolved. The reconstructions in Figure~\ref{faststack_coverage_fig}  were made using $\alpha=0.05$, however the appearance of the reconstructed images is not overly sensitive to the value of $\alpha$ chosen.  Figure~\ref{alpha_comparison_fig} shows the faststack reconstructions for two different values of $\alpha$, for the region with the average coverage of transits. The left column shows reconstructions for $\alpha=0.05$ and the right column for $\alpha=0.45$, both of these values, and those in between, produce good reconstructions in terms of identifying the locations of the secondary sources present, and the absence of artefacts.  If  $\alpha$ is increased further the images begin to show the artefacts due to the long, narrow samples, as the initial estimates become dominated by the larger samples values due to transits of the primary.  The top row of Figure~\ref{alpha_comparison_fig} shows the reconstructions for data used in Figure~\ref{faststack_coverage_fig}, for the average coverage region. The second row shows reconstructions, for fainter secondary sources, the magnitude difference between the primary and the secondaries being increased to one, thereafter for every subsequent row the magnitude difference between the primary and the secondaries is increased by one, reaching a magnitude difference of five, by the last row. Additionally, for all image reconstructions, bar those in the first row, a $0.2^{\prime \prime}$ radius disc centred on the primary has been masked out, in order to reduce the range of the colour scale, to aid in the visualisation of the secondaries.  What is noticeable in these images, is that for lower values of $\alpha$, the appearance of all the sources is more compact, and for secondaries that have a large magnitude difference from the primary this may result in them being more easily distinguishable. 

\begin{figure}
\includegraphics[width=0.505\textwidth]{faststack_a1_lowalpha}\includegraphics[width=0.505\textwidth]{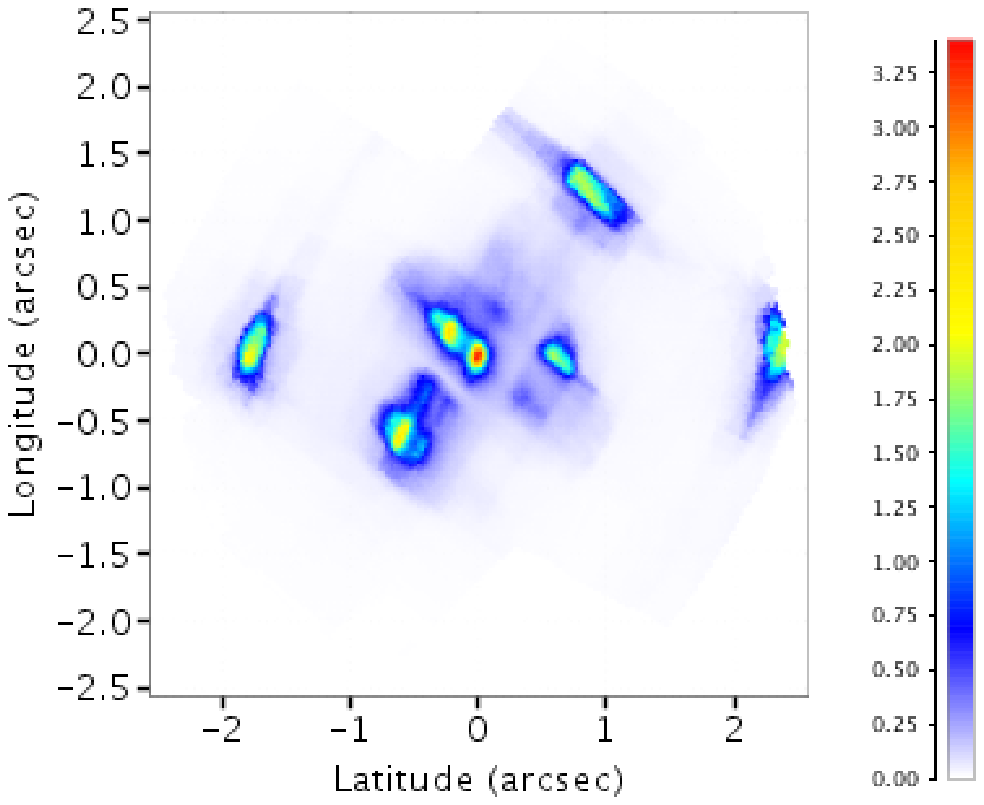}
\caption{This plot shows the faststack reconstruction, corresponding to the low coverage region shown in the top panel of Figure~\ref{coverage_fig}, with the secondaries half a magnitude fainter than the primary. On the left, $\alpha=0.05$ and on the right, $\alpha=0.45$. In this case when the higher value of $\alpha$ is used, the presence of artefacts due to the long, narrow samples are clearly noticeable.}
\label{alpha_lowcoverage_comparison_fig}
\end{figure}

The range of acceptable values for $\alpha$ reduces in regions of non-uniform coverage, and especially those with low coverage, where the higher value of $\alpha$ used in Figure~\ref{faststack_coverage_fig} already results in images with noticeable artefacts. This may be seen in Figure~\ref{alpha_lowcoverage_comparison_fig}, where the image reconstruction has been performed using the data from the low coverage region shown in Figure~\ref{faststack_coverage_fig} for $\alpha=0.05$ on the left and $\alpha=0.45$ on the right. A sensible approach to choosing $\alpha$ therefore would be to choose a low value, in order to ensure good reconstructions in regions of low and non-uniform coverage.

The difficulties in performing an image reconstruction through thresholding, are also apparent from Figure~\ref{alpha_lowcoverage_comparison_fig} as this shows that for certain pixels in the image the initial estimates for their values are already contaminated by flux from the primary long before even the median value contributing to the pixel is included. One could think to use the initial estimate alone and to discard the remaining data, effectively turning $\alpha$ into a parameter determining the level of thresholding. This produces visually similar images, but the values in the reconstructed image pixels depend much more heavily on the value of $\alpha$ chosen; whereas when the weighting scheme described here is used the reconstructed images for different values of $\alpha$ are are much more consistent. 

\section{Comparison with the Drizzle algorithm}
\label{drizzle}

\begin{figure}
\includegraphics[width=0.505\textwidth]{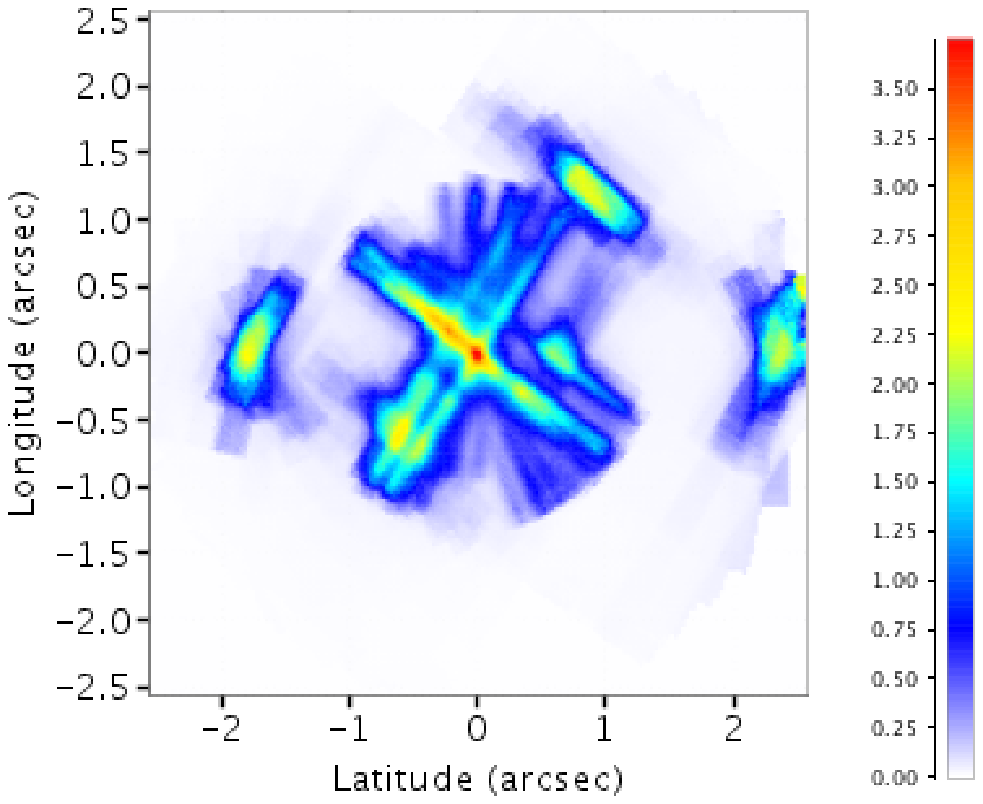}\includegraphics[width=0.505\textwidth]{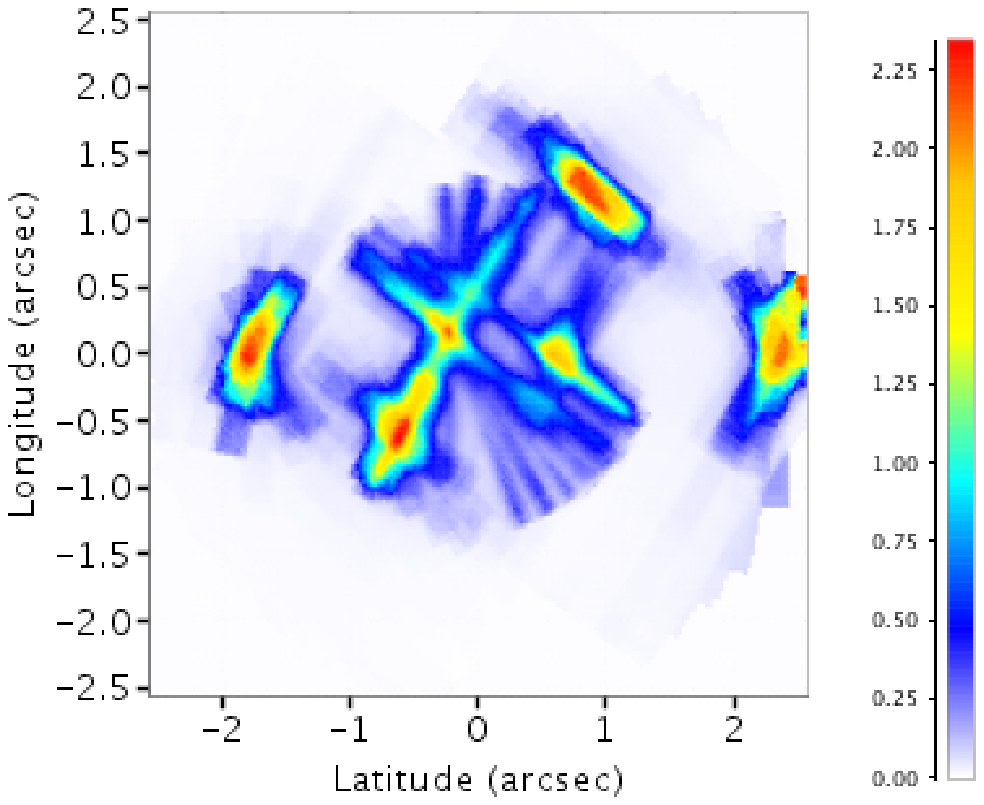}
\includegraphics[width=0.505\textwidth]{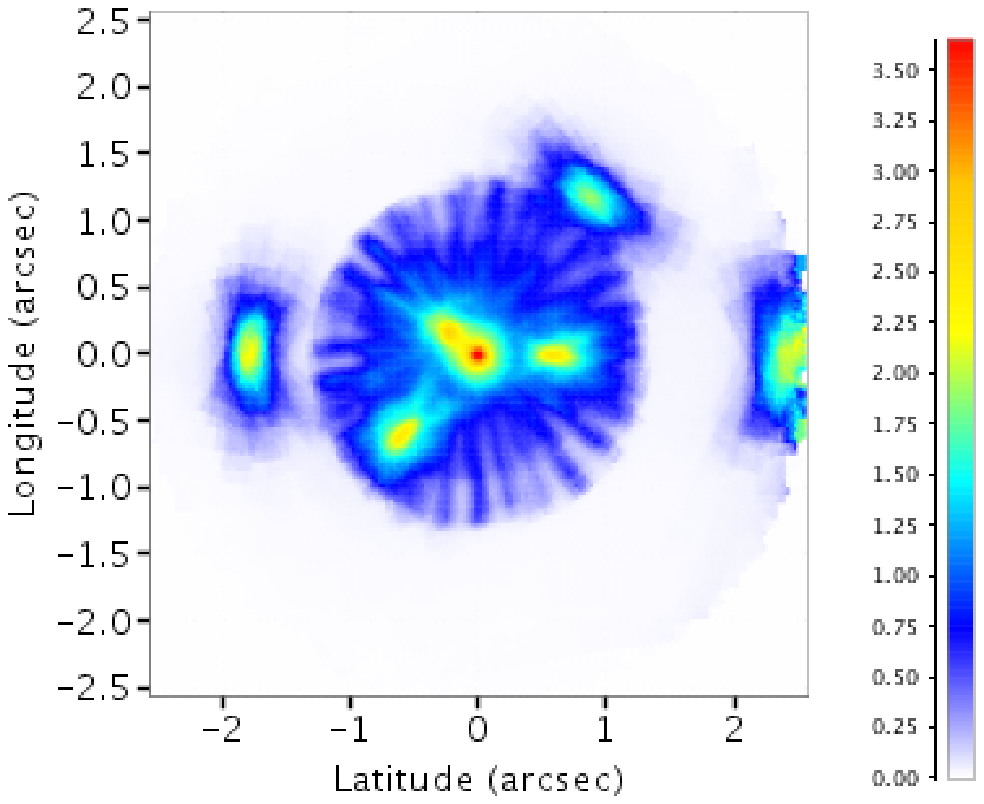}\includegraphics[width=0.505\textwidth]{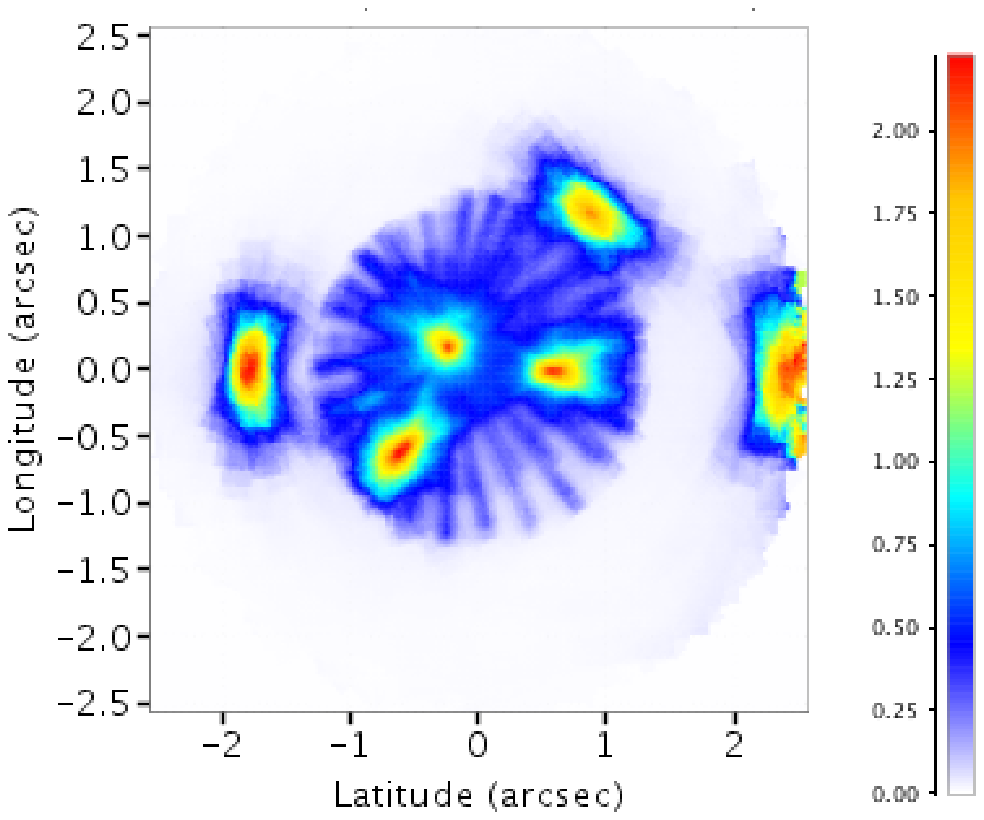}
\includegraphics[width=0.505\textwidth]{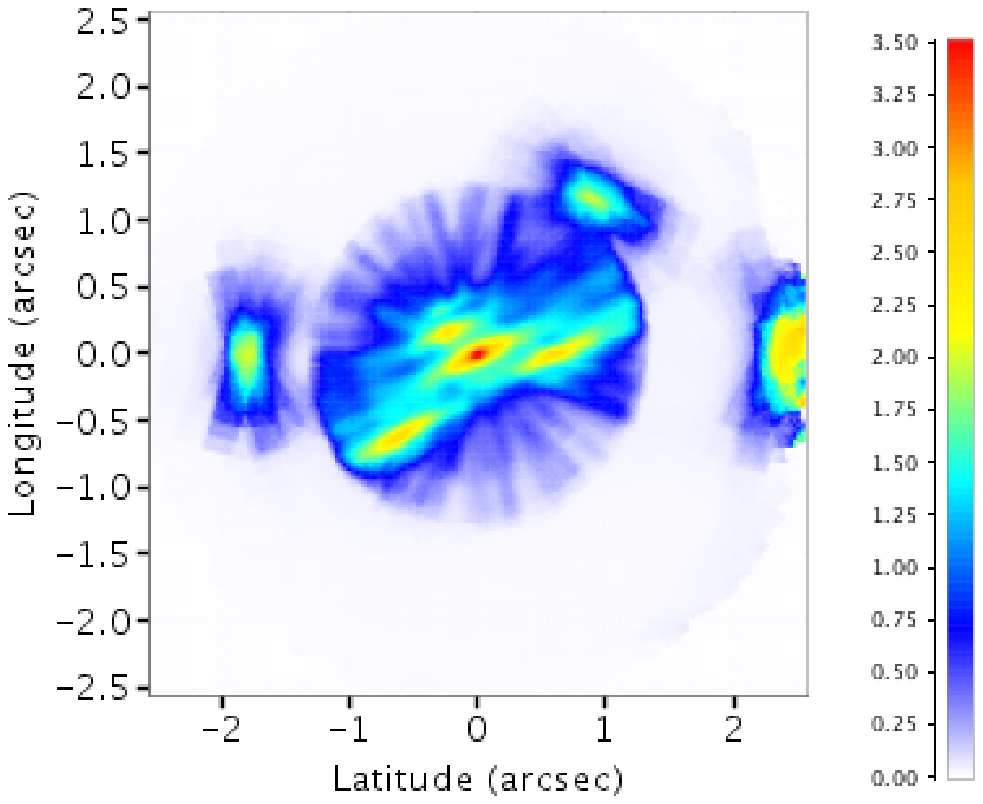}\includegraphics[width=0.505\textwidth]{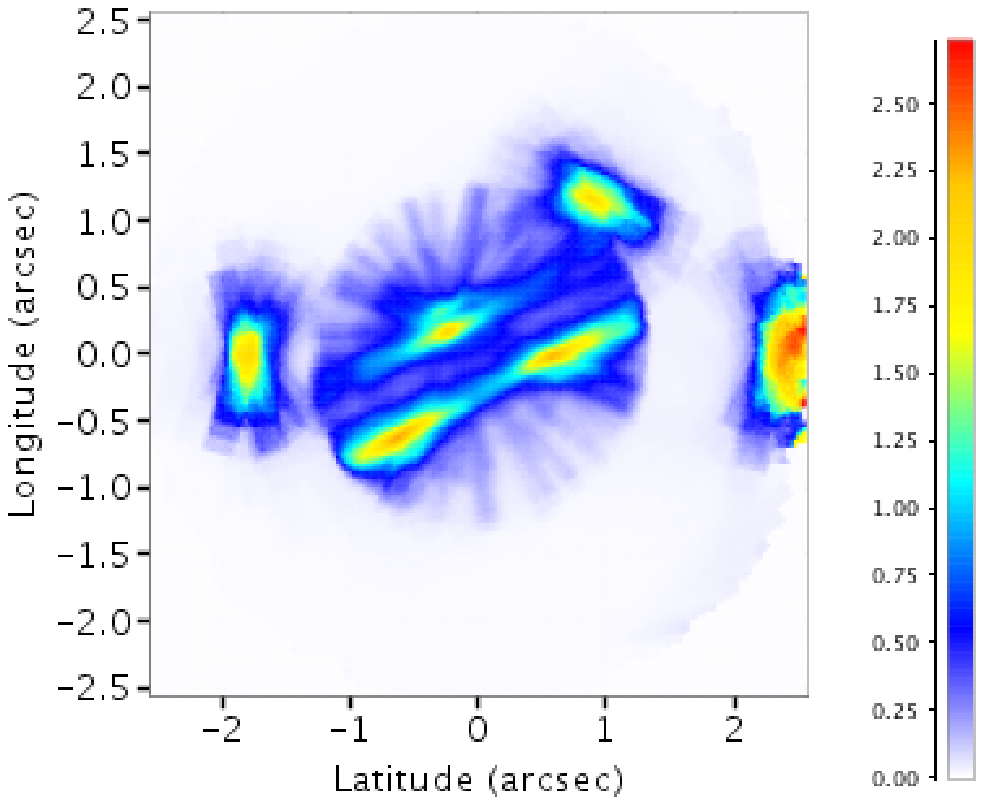}
\caption{This figure shows the images reconstructed by the drizzle method in the regions corresponding to the coverages shown in Figure~\ref{coverage_fig}. The reconstructions use the SM, AF2, 5 and 8 data,  on the left and on the right using the same data, but subtracting the contribution of the primary with using the PSF and magnitude used to produce the simulations.}
\label{drizzle_coverage_fig}
\end{figure}

\begin{figure}
\includegraphics[width=0.505\textwidth]{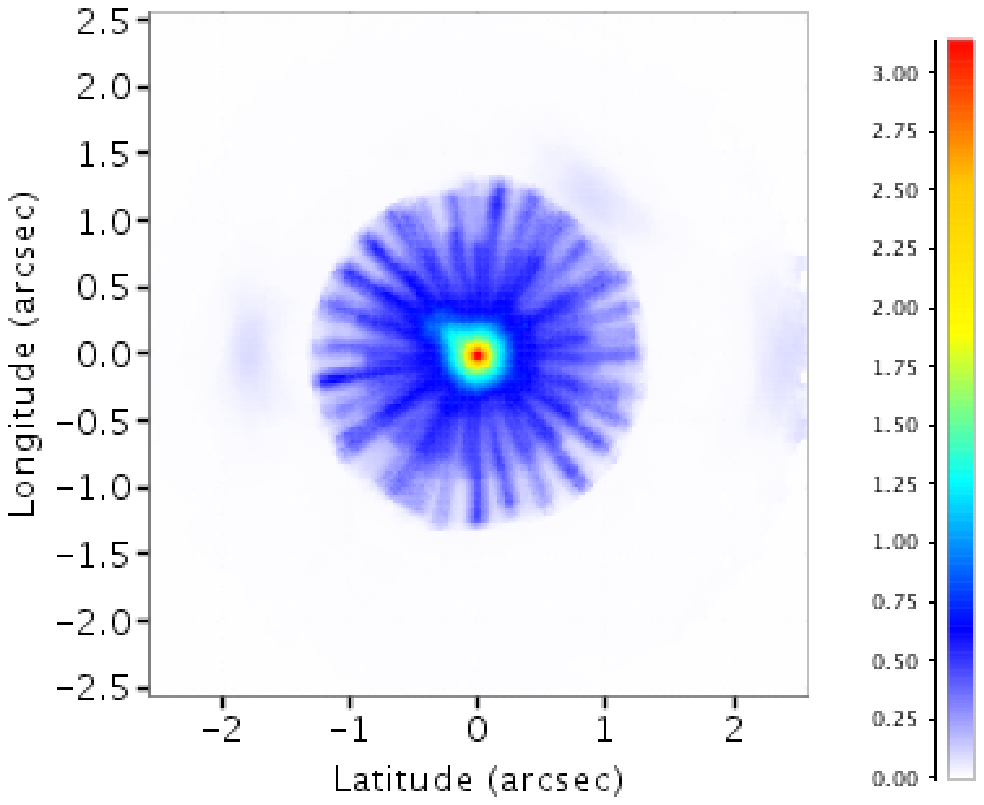}\includegraphics[width=0.505\textwidth]{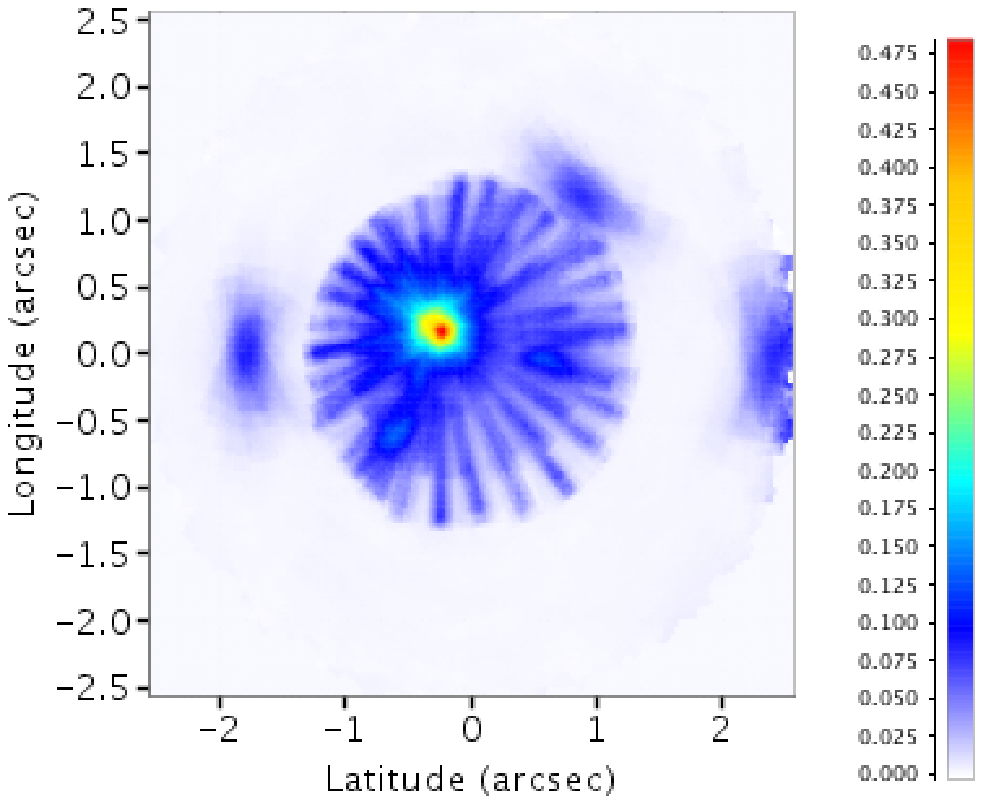}
\includegraphics[width=0.505\textwidth]{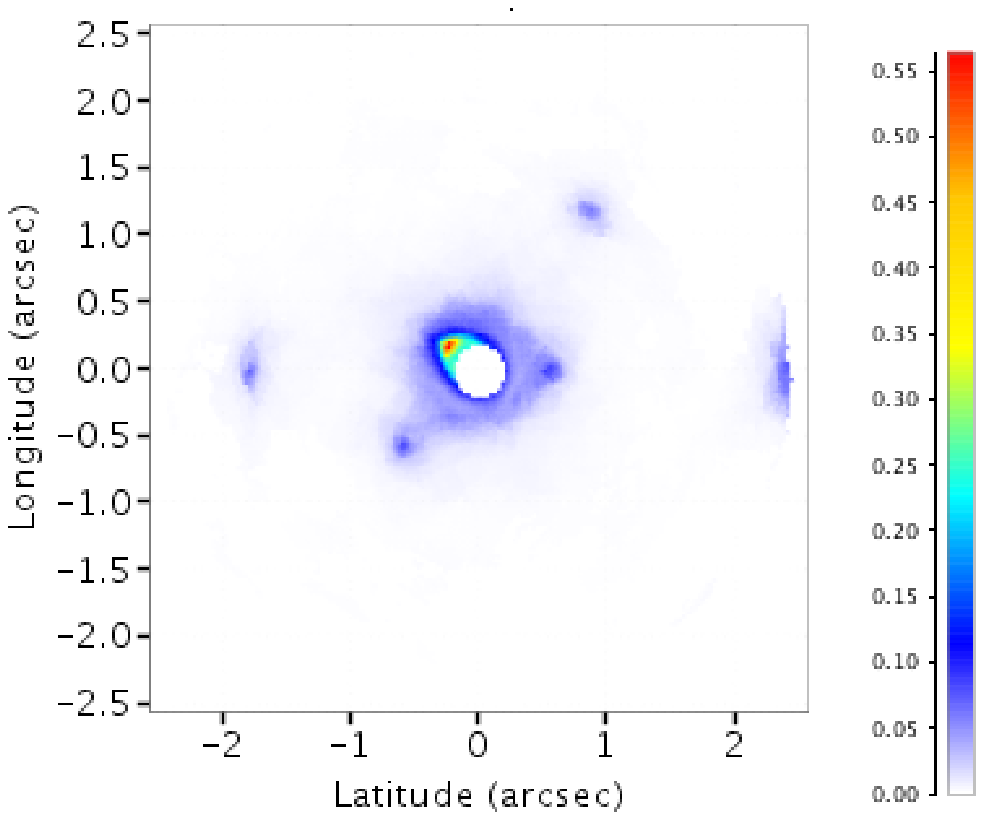}\includegraphics[width=0.505\textwidth]{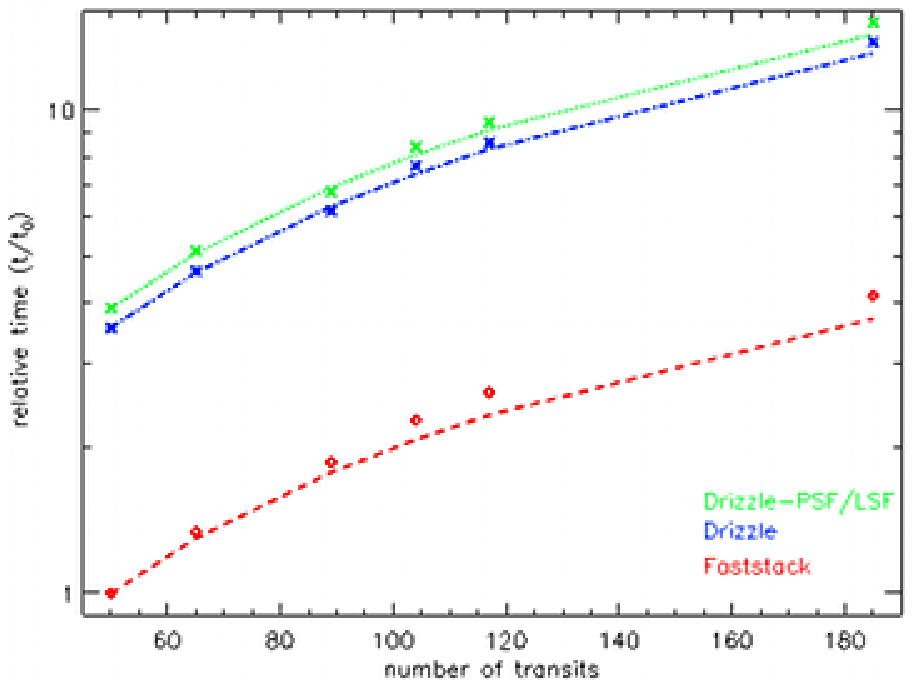}
\caption{A comparison of reconstructions using drizzle, top left, drizzle subtracting the contribution of the primary prior to reconstruction, top right, and faststack, bottom left, with the primary masked. The primary has a magnitude of 18, source B1 has a magnitude of 20,  the remaining secondaries have magnitudes of 22,  and the distribution and number of transits correspond to the average coverage. Here it is possible to see that the artefacts due to a secondary source may become important; leading to the need to iterate the image reconstruction if  sources, B2 and B3 are to be discovered. The final panel, bottom right, shows the relative timings between all three image reconstructions displayed as a function of the number of transits of the primary.}
\label{drizzle_faststack_comp_fig}
\end{figure}

Image reconstruction using the drizzle algorithm, \cite{fruchter02}, has been investigated in the case of {\it Gaia} before by \cite{nurmi05}. This used a previous expectation for the windowing and binning of the data, as well as assuming perfect knowledge of the line spread function, LSF,  so it is worth revisiting this in order to demonstrate the difficulties drizzle has with the long, narrow samples of the intermediate and faint source AF windows. Figure~\ref{drizzle_coverage_fig} shows the reconstructions made using drizzle, for data simulated in each of the coverage regions shown in Figure~\ref{coverage_fig}, and on which the faststack reconstructions in Figure~\ref{faststack_coverage_fig} were performed.  The left-hand side of Figure~\ref{faststack_coverage_fig} corresponds directly to the left-hand side of Figure~\ref{drizzle_coverage_fig}, in that exactly the same data is used.  Figure~\ref{drizzle_coverage_fig} shows the problems created by the 1-dimensional AF windows  in the form of linear artefacts aligned with regions of increased coverage, top and bottom plots, and in the centre, as an increase in the level of the background for incidences of relatively uniform coverage. These linear artefacts in the reconstructed images are due to the long, narrow window-samples and the inability to localise this flux.  These artefacts result predominantly from window-samples which cross the primary. The approach,  therefore,  used by \cite{nurmi05}  to deal with this issue was to subtract the flux due to the primary from the window data prior to the reconstruction.  This approach is shown on the right-hand side of Figure~\ref{drizzle_coverage_fig},  where the contribution of the primary source has been removed from each window used in the reconstruction, using both the magnitude of the primary and the PSF and LSFs  used in simulating this data.  This side of the figure shows that the artefacts due to the secondary sources may also be important. This is illustrated further in Figure~\ref{drizzle_faststack_comp_fig}, where not all the secondaries have the same magnitude. In this figure the source, B1 has a magnitude of 20, the remaining secondaries have magnitudes of 22 and the primary as always has a magnitude of 18. The top left panel shows the drizzle reconstruction, and the top right the drizzle reconstruction after the subtraction of the primary, here we can see how the source B1 now causes similar problems to the primary source in the standard drizzle reconstruction. To recover the sources, B2 and B3 it is likely that B1 must also be subtracted from the windows prior to reconstruction making this an iterative procedure.

The bottom left panel in Figure~\ref{drizzle_faststack_comp_fig} shows the faststack reconstruction for the same data-set for comparison, masking out the primary with a disc of radius $0.2^{\prime\prime}$, here we can see how all the sources could be discovered in a single pass. The bottom right panel shows the relative timings between drizzle and faststack as a function of the number of transits of the source. As expected both methods scale linearly, as indicated by the lines in the figure, with the number of transits. The timings for drizzle when subtracting the primary source take fractionally longer than the standard drizzle, since the contribution of the primary to each window must be calculated before it can be removed. However, should a secondary source need to be subtracted then this would double the time required, since first it must be discovered in the reconstructed image.

 Additional concerns with this approach are that the PSF, LSFs, and magnitudes of the sources to be subtracted must be well known if artefacts due to this subtraction are not to be introduced.  There is also the issue of the treatment of extended sources in this scheme, since if the primary is assumed to be point-like when subtracted this would severely complicate the analysis of the reconstructed image.
 
\section{Discussion}
\label{discussion}

An image reconstruction method has been demonstrated, which can successfully use  the 1-dimensional data expected from {\it Gaia}. This method makes no assumptions about the LSF or PSF, unlike the previous image reconstruction methods investigated for use with {\it Gaia} data (\cite{nurmi05}, \cite{lindstroem06}, \cite{mary06}).  This is advantageous as uncertainties in these quantities will not effect the reconstructed image, and with the severe constrains on the CPU time available per source given the size of the {\it Gaia} catalogue, the reduction in the amount of data required to be read per image reconstruction may result in an appreciable amount of CPU time, as a different PSF/LSF would be required for each strip and row of the CCD; source colour, and potentially for each individual transit due to  the effects of radiation damage.  

The approach used by \cite{nurmi05} of subtracting the contribution of the primary source to improve the magnitude difference to which secondaries may be detected in the drizzle reconstructions, is unlikely to be feasible for use with the {\it Gaia} data for the various reasons discussed above,  which all reduce to the degree of complexity required and hence the amount data processing needed.  With and without the subtraction of the primary there are artefacts in the drizzle images, due to the 1-dimensional data which increase the difficulty of the analysis of the reconstructed image. Faststack, however, due to its weighting scheme, successfully reconstructs images using this 1-dimensional data; this greatly simplifies the analysis of the resultant images.  The  weighting scheme adopted for faststack is naturally robust against hot pixels, which may be caused by solar proton or cosmic-ray hits; the sample to which the hot pixel contributes would naturally be weighted down. The other image reconstruction methods would need some form of preprocessing to remove the effected data prior to reconstruction, adding to their CPU budget.

After the reconstructed image has been analysed and if it is found to contain any additional sources, the next stage of the processing is performed. It is at this stage that the disturbing effects of the secondary sources are accounted for and corrections to the astrometric and photometric parameters of the primary source are made; as well as the addition of the newly-discovered sources to the catalogue.  This stage of the processing refines the image parameters for all sources found in the reconstructed image. It takes as initial values for the image parameters of position and magnitude, those values extracted from the reconstructed image, together with the original transit data and PSF/LSF data. The analysis proceeds via a least squares approach as explained in  \cite{vanLeeuwen09}; the primary difference being that in addition to position and magnitude, the proper motions and differential parallaxes are also solved for. 

The importance of the image reconstruction in the overall quality of the final catalogue is apparent, as it is this stage which controls whether a secondary source is discovered or not; and should any spurious detections occur at this stage, then at worst it results in a spurious source in the final catalogue or at best it wastes precious CPU time in the image parameters analysis where it is rejected as a spurious detection.

Faststack  is currently  the most suitable method for performing image reconstructions using the {\it Gaia} data, for the purpose of discovering secondary sources in the vicinity of the primary sources. This is due to its lack of complexity, hence speed, and ability to produce images which appear free from artefacts, and hence should not require a complicated, CPU intensive  algorithm,  to extract the locations of the secondaries from the image without producing spurious detections. 

\begin{acknowledgements}
Simulated data provided by the Simulation Unit (CU2) of the Gaia Data Processing Analysis Consortium (DPAC) have been used to complete this work. The simulations have been done at CNES (Centre national d'\'{e}tudes spatiales). They are gratefully acknowledged for this contribution.
\end{acknowledgements}

\bibliographystyle{aps-nameyear}      
\bibliography{faststack}   

\end{document}